%% file: mage-sample.tex
\documentclass[iop,apj,revtex4]{emulateapj}
\input{definitions} 
\usepackage{pdfpages, subfloat, xcolor} 
\newcommand{\samplesize}{$N=15$}
\newcommand{\avgR}{$R=3300$}
\newcommand{\coshst}{COS/\textit{HST}}

\slugcomment{Submitted to ApJ 26 June 2017.  Re-resubmitted 31 Oct 2017.}

\shorttitle{Rest-frame UV spectra at $z\sim2$}
\shortauthors{Rigby \etal}
\begin{document}


\title{The Magellan Evolution of Galaxies Spectroscopic and 
Ultraviolet Reference Atlas (\megasaura) I:  The Sample and the Spectra}
\author{J.~R.~Rigby\altaffilmark{1}, 
M.~B.~Bayliss\altaffilmark{2}, 
K.~Sharon\altaffilmark{3}, 
M.~D.~Gladders\altaffilmark{4,5},
J.~Chisholm\altaffilmark{6},
H.~Dahle\altaffilmark{7},  
T.~Johnson\altaffilmark{3}, 
R.~Paterno-Mahler\altaffilmark{3}, 
E.~Wuyts\altaffilmark{8}, 
and D.~D.~Kelson\altaffilmark{9}}
\altaffiltext{1}{Astrophysics Science Division, 
           Goddard Space Flight Center, 8800 Greenbelt Rd., Greenbelt, MD 20771 USA}
\altaffiltext{2}{MIT Kavli Institute for Astrophysics and Space Research, 
          77 Massachusetts Ave., Cambridge, MA 02139 USA}
\altaffiltext{3}{Department of Astronomy, University of Michigan, 
          500 Church St., Ann Arbor, MI 48109 USA}
\altaffiltext{4}{Department of Astronomy \& Astrophysics, University of
           Chicago, 5640 S. Ellis Ave., Chicago, IL 60637 USA}
\altaffiltext{5}{Kavli Institute for Cosmological Physics, University of
          Chicago, 5640 South Ellis Ave., Chicago, IL 60637 USA}
\altaffiltext{6}{Department of Astronomy, University of Geneva, Geneva, Switzerland}
\altaffiltext{7}{Institute of Theoretical Astrophysics, University of Oslo, 
              P.O. Box 1029, Blindern, NO-0315 Oslo, Norway}
\altaffiltext{8} {ArmenTeKort, Antwerp, Belgium}
\altaffiltext{9}{The Observatories, Carnegie Institution for Science, 813 Santa Barbara St., 
         Pasadena, CA 91101 USA}
\email{Jane.Rigby@nasa.gov}

\begin{abstract}
We introduce Project \megasaura: \megasauralong.  
\megasaura\ comprises medium-resolution, rest-frame ultraviolet spectroscopy 
of \samplesize\ bright gravitationally lensed galaxies at redshifts of 1.68$<$z$<$3.6, 
obtained with the MagE spectrograph on the Magellan telescopes. 
The spectra cover the observed-frame wavelength range
$3200 < \lambda_o <  8280$~\AA ; 
the average spectral resolving power is \avgR. 
The median spectrum has a signal-to-noise ratio of $SNR=21$ per resolution element at 5000~\AA . 
As such, the \megasaura\ spectra have superior signal-to-noise-ratio and wavelength coverage
compared to what \coshst\ provides for starburst galaxies in the local universe.  
This paper describes the sample, the observations, and the data reduction.  
We compare the measured redshifts for the stars, the ionized gas as traced by nebular lines, 
and the neutral gas as traced by absorption lines; we find 
the expected bulk outflow of the neutral gas, and no systemic offset between 
the redshifts measured from nebular lines and the redshifts measured from the stellar continuum. 
We provide the \megasaura\ spectra to the astronomical community 
through a data release.
\end{abstract}


\section{Introduction}\label{sec:intro}

A major driver for the construction of 20 to 30~m optical telescopes 
is the desire to obtain rest-frame 
ultraviolet (UV) spectral diagnostics for large numbers
of  galaxies over most of cosmic time  \citep{Skidmore:2015te}.
Rest-frame ultraviolet spectra  constrain the hot stellar populations in galaxies 
via P~Cygni stellar wind profiles, 
and constrain the  nucleosynthetic buildup of heavy elements through 
nebular emission lines (measuring metallicity in the H~II regions) and
weak photospheric absorption lines (measuring metallicity in the 
hot stars, c.f. \citealt{Pettini:2000dy}.)

Rest-frame UV spectroscopy also reveals galactic-scale winds: 
the mass outflow rate, the velocity structure, ionization 
structure, metallicity, and abundance pattern of the outflowing gas 
\citep{Pettini:2002gb, Quider:2009jd, Quider:2010cg}. 
These winds are the famous ``galactic feedback'' 
\citep{Heckman:1990vt, Heckman:2000du, Veilleux:2005ec, Martin:2005kx}%
--  they drive metals into the intergalactic medium, and they may shut down 
future star formation.  In addition, these wavelengths feature a number
of emission lines, most of them weak, that act as spectral diagnostics
of the electron temperature, ionization parameter, and abundance pattern  \citep{Bayliss:2014ib}

However, with current telescopes, it has been extremely challenging to obtain, 
for individual galaxies, both the high signal-to-noise ratios 
and moderate spectral resolution ($R\sim 3000$) 
required to make full use of the rest-frame ultraviolet diagnostics.  
For example, the high-quality,  
R$=$750 spectrum analyzed by \citet{Erb:2010iy} required 15~hr of integration on Keck; 
still, this is lower spectral resolution than is ideal for narrow features 
like photospheric absorption lines.
An alternate approach has been to stack the spectra of field (unlensed) galaxies
(for example, \citealt{Shapley:2003gd, Jones:2012kn, Steidel:2016wu, Zhu:2015kv}).
The main risks of stacking are that it can be influenced by outliers, and 
that it washes out heterogeneity within the sample being stacked.  
In addition, stacks published to date have had lower spectral resolving
power ($R=560$, 660, 1400, and 2000 for the four citations above)
than is ideal.

A third approach, the one adopted in this paper, is to take
advantage of rare cases where galaxies have been highly magnified 
by gravitational lensing.  For the brightest lensed galaxies, 
medium--resolution rest-frame ultraviolet spectroscopy
can be obtained in reasonable integration times with 6-10~m telescopes.  
By targeting such bright lensed galaxies, spectra can be obtained
with current telescopes that can independently 
measure metallicities for the hot stars, the interstellar medium, and 
the H~II regions within individual galaxies.  Such multiple 
constraints enable vital cross-checks of the bright rest-optical 
metallicity diagnostics \citep{Steidel:2016wu}, 
whose empirical calibrations should evolve with redshift \citep{Kewley:2008be}.

To date, only five strongly lensed $z\sim$2--3 galaxies have been dissected
with high-quality rest-frame UV spectroscopy:
cB58 \citep{Pettini:2000dy,Pettini:2002gb}; 
FORJ0332$-$3557 \citep{Cabanac:2008iw}; 
the Cosmic Horseshoe \citep{Quider:2009jd}; 
the Cosmic Eye \citep{Quider:2010cg};
and the Eight O'clock arc \citep{DessaugesZavadsky:2011cb}.
Each galaxy measured to date has very different outflow 
properties \citep{Quider:2010cg}; clearly, the published sample size is simply
too small to characterize outflows
and massive star content at z$\sim$2--3.

To directly address this problem, we have conducted Project \megasaura:   
\megasauralong .
\megasaura\ has obtained rest-UV spectra of \samplesize\ bright lensed galaxies, 
taken with the MagE instrument on the Magellan telescopes.   
This is the sample and data paper for \megasaura, in which we describe
the observations and the spectra.  We also measure the velocity offsets 
between the nebulae, stars, and interstellar media  within these lensed galaxies. 

\section{Methods}\label{sec:methods}

\subsection{The sample}\label{sec:sample}
The \megasaura\ sample is comprised of 
\samplesize\ bright strongly--lensed galaxies, 
spanning the redshift range 1.68$<$z$<$3.6, including
many of the brightest lensed galaxies known. 
The criterea for target selection were: 
highly-magnified lensed galaxies, 
as bright as possible in the observed g-band, 
with an image plane configuration such that a significant portion 
of the total emission would fit into a 10\arcsec\ MageE slit with
room for sky on both ends of the slit, 
with a redshift such that 
both Ly~$\alpha$ and C~IV are observable by MagE, 
and at a sufficiently southern declination to be observable by Magellan.

Most of the \megasaura\ targets are behind cluster--scale lenses 
that were discovered through two related surveys:
SDSS Giant Arcs Survey 1 (SGAS1, Gladders \etal\ in prep.), 
and SDSS Giant Arcs Survey 2 (SGAS2, Dahle \etal\ in prep.)
SGAS1 has yielded 217 lenses, with a 
median giant arc redshift of z$=$ 2 \citep{Bayliss:2011fr,Bayliss:2012dy};
SGAS2 expanded the search space within the Sloan Digital Sky Survey (SDSS), 
resulting in the discovery of 62 additional lenses.

We also included three targets from other samples: 
the bright lensed galaxy \rcsohthree  \citep{Wuyts:2010gy}, 
which was discovered through the Red-sequence Cluster Survey-2 (RCS-2, \citealt{Gilbank:2011dn}); 
and two bright galaxy-scale lenses:   
the Cosmic Eye \citep{Smail:2007bq} and the Cosmic Horseshoe \citep{Belokurov:2007bv}.

Table~\ref{tab:sample} lists the discovery paper for each arc;
several targets were independently discovered by multiple groups.
Table~\ref{tab:metallicities} lists oxygen abundances as known for the sample;
7 galaxies have measurements or constraints.
The measurements range from $<25\%$ of solar  to $81\%$ of solar, with a
median of $37\%$ of solar (where solar is from \citealt{Asplund:2009eu}.)
Several of the \megasaura\ galaxies have redshifts
such that the rest-frame optical oxygen abundance
diagnostics are not accessible from the ground;
access to these diagnostic lines was not a selection criterion.
Oxygen abundances for other SGAS galaxies,
not in the \megasaura\ sample, were published in \citet{Wuyts:2012gb};
the median and quartile ranges are $66^{+27}_{-36}\%$ of solar.

The \megasaura\ sources are exceptionally bright: the brightest 
has $g_{AB}=19.15$ \citep{Wuyts:2010gy}, and most have $g_{AB} \sim20$-–21 
(\citealt{Koester:2010ky}; Fig.~5 of \citealt{Wuyts:2012ej}).
For comparison, individual field galaxies that went into the composite spectrum 
of \citet{Shapley:2003gd}  have $g_{AB} \sim 24.7$  \citep{Erb:2006fh}.  
The \megasaura\ targets are many of the brightest lensed sources 
selected from the Sloan Digital Sky Survey.  

Table~\ref{tab:sample} lists the targets, coordinates, and total integration times.
Figure~\ref{fig:where_pointed} and Figure~\ref{fig:where_pointed_rc0327} 
contain findercharts for the \megasaura\ sample.

Portions of the \megasaura\ spectra have been published previously.  
\citet{Koester:2010ky} published smoothed spectra for \stwelve\ and \sfifteen.  
\citet{Rigby:2014hq} pointed out the lack of correlation between the Mg~II emission 
and Ly~$\alpha$ emission in five of these galaxies.
\citet{Bayliss:2014ib} published the MagE spectrum of the highest redshift galaxy 
in the sample,  SGAS~J105039.6$+$001730  at z$=$3.6252.
\citet{Rigby:2015jy} analyzed  [C~III]~1907,  C~III] 1909~\AA\ emission for the sample.
\citet{Bordoloi:2016js} analyzed variation in the Mg~II and Fe~II emission and absorption 
among four star-forming knots within lensed galaxy \rcsohthree .

\subsection{Spectroscopic observations}\label{sec:obs}
Observations were obtained with The Magellan Echellette (MagE) spectrograph 
\citep{Marshall:2008bsa} 
on the twin 6.5~m Magellan telescopes at the Las Campanas Observatory in Chile. 
The majority of observations were conducted with MagE mounted on the Clay telescope;
some observations were made after October 2015, when MagE was moved to the Baade telescope. 
Targets were acquired by blind offsets from bright nearby stars; target acquisition was verified via the slit-viewing
guider camera.  MagE has eight available slits, all 10\arcsec\ in length.  We used slit widths of 
0.7, 0.85, 1.0, 1.5, and 2\arcsec, with the slit width chosen to match the atmospheric 
seeing and the compactness of the emission.  

For two galaxies, \rcsohthree\ and \sfifteen,  we obtained spectra
of multiple physically distinct regions, as noted in Table~\ref{tab:sample} and 
on the findercharts in Figure~\ref{fig:where_pointed} and Figure~\ref{fig:where_pointed_rc0327}.

The broad wavelength coverage of MagE makes it important to keep the slit position 
angle close to the parallactic angle, especially when the airmass is significant 
\citep{Filippenko:1982br}.  This was done by advancing the slit angle 
to where the parallactic angle was expected 30 minutes hence,  
acquiring the object,  exposing (typically for one hour),  and repeating. 
A few exceptions to this rule were made to keep contaminating objects out of the slit.  
The slit position angle and parallactic angle are noted in Table~\ref{tab:obslog}.  

Because these lensed galaxies are highly spatially extended, and because we maintained the 
parallactic angle, we generally obtained spectra for only a portion of each lensed galaxy.
This spatial coverage is 
illustrated in Figure~\ref{fig:where_pointed} and Figure~\ref{fig:where_pointed_rc0327}, 
which show the slit position angles obtained for each galaxy in the sample.

We obtained calibrations nightly, following the recommended
sequences in the MagE manual. 
Internal Th-Ar lamps were obtained for wavelength calibration.  
Xe-flash lamp exposures, using the same slit as the science observations, 
were obtained to define where the spectral orders fall on the detector.

The wide wavelength coverage of MagE makes
it a challenging instrument to flat-field.  
Our strategy was to take three kinds of flat-field calibrations:
calibrations for the bluest two orders, for the other blue orders, and for the red orders.
To obtain a high quality flat-field for the bluest two orders, we
obtained a sequence of calibration frames (``very blue flats'') using 
the internal Xe-flash lamp,  long (100~s) integration times,
and  with the 5\arcsec\ slit and the instrument defocused to blur out the
broad Xe emission lines.    These frames have good counts in the 
two bluest two orders, and saturate the others.
To calibrate the remaining blue orders, we obtained 
shorter (10--15~s) Xe-flash exposures (``blue flats''), 
again with the 5\arcsec\ slit and the instrument defocused.
To calibrate the redder orders, we obtained calibration frames
(``red flats'') using a quartz lamp mounted in the secondary cage, 
and the same slit as the science observations. 

Nightly we observed spectrophotometric standard stars, chosen from the 
Nearby Supernova Factory Spectroscopy Standards 
list.\footnote{\url{http://snfactory.lbl.gov/snf/snf-specstars.html}}

\subsection{Spectroscopic pipeline processing}\label{sec:dataredux1}
For each night, calibrations were generated and the data were processed using the
MagE pipeline, which is part of the Carnegie Python 
Distribution\footnote{http://code.obs.carnegiescience.edu}. 
Here we summarize what the pipeline does.

Data frames first were subtracted of their bias levels, using the overscan regions. 
Second, Th Ar lamp frames were used to define the regions of the chip 
covered by the spectral orders. 
The inter-order regions were then interpolated to generate a non-parametric model
of the scattered light.

Next, two flat-fields were generated: the blue flat frames and very blue flat
frames were masked of their saturated regions and combined using a weighted average, 
to produce a blue flat, with  high signal-to-noise in the blue, but saturated in the red.
The red flat frames were combined in the same way to produce a red flat,
that is unsaturated in the red, but has low signal-to-noise in the blue.
The blue flat and the red flat were then combined by masking out saturated regions, and 
combining with a weighted average to create a flat field.
This flat-field was then used to flat-field the individual spectral exposures.

Individual spectra were then sky subtracted.  The sky emission was modeled
using routines described in \citet{Kelson:2003iz}, 
but with explicit masking of pixels that are contaminated by the target.  
The requisite rectification transformations to account for line curvature
were derived using  routines described in \citet{Kelson:2000iy} 
that analyze the tracings of cross correlations of image sections in the comparison 
Th Ar lamp frames.

After sky subtraction, one-dimensional spectra for the
objects were extracted using a multi-exposure optimal extraction
algorithm\footnote{Described at http://code.obs.carnegiescience.edu/}, 
 in which the spatial profile
of the object in each exposure is characterized by a Gauss-Hermite
decomposition with moments that are polynomial functions of order
number and position across the detector. With the spatial profile of
the object characterized in this manner for each exposure, a single
univariate B-spline \citep{Dierckx:1983fe} representation of the object
spectrum can derived that, in a least-squares sense, best represents
all of the available data simultaneously.

The output of the pipeline is an extracted, one-dimensional, 
wavelength-calibrated spectrum for each echelle order. 

\subsection{Spectroscopic post-pipeline processing}\label{sec:dataredux2}
In order to flux the spectra, 
we computed the sensitivity function for each night using the 
IRAF\footnote{IRAF is distributed by the National Optical Astronomy Observatory, 
which is operated by the Association of Universities for Research in Astronomy (AURA) 
under a cooperative agreement with the National Science Foundation.}  
tools {\bf standard} and {\bf sensfunc} in the  {\bf noao.onedspec} package.  The
sensitivity functions for each standard star observation typically show a spread
in normalization of up to $20$--$30\%$ peak-to-peak in the red, and up
to $40\%$ in the bluest orders; this is consistent with expected
seeing variations, as gauged by the Magellan seeing monitor.  
We scaled the sensitivity functions to the star
with the highest throughput to create a composite sensitivity
function.  To flux calibrate, these sensitivity functions were applied
to each spectrum using the IRAF tool {\bf onedspec.calibrate}.  

After fluxing, we combined the 
overlapping echelle orders, using a weighted average, weighted by the inverse variance, 
to make a continuous spectrum for each target for each night. 
Each  spectrum was then corrected to vacuum barycentric wavelength.  
For each galaxy, the spectra from multiple nights were combined with a 
weighted average, weighted by the inverse variance.  
For the two galaxies where we obtained spectra of multiple distinct physical regions, 
as noted in Table~\ref{tab:sample}, we separately combine the spectra of each
distinct region.

The output spectra have vacuum barycentric wavelength in \AA\ and 
flux density f$_{\nu}$ in \cgsfnu.

There are significant telluric absorption features redward of
6400~\AA, primarily the A-band of molecular
oxygen, which covers 7590~\AA\ to 7660~\AA\ \citep{Wark:1965gu},  
and the B-band (6860--6890~\AA).
For observing runs where the featureless standard star EG~131 was observed, we 
used that star and the IRAF tool {\bf noao.onedspec.telluric} to correct
those telluric features in the two spectral orders that are centered at 
6730 and 7570~\AA.

The brightest night sky lines,  [O I] 5577\AA\ and [O I]~6300\AA,
frequently saturate individual exposures, which results in artifacts at those wavelengths. 

The MagE spectrograph is extremely sensitive in the blue, with sensitivity
down to 3000~\AA.  However, the 
Earth's atmospheric transmission at 3000~\AA\ is only $0.6\%$, 
as compared to $17.7\%$ at 3200~\AA\ and $28.7\%$ at 3400~\AA\ 
(\citealt{Cox:2000ua} Table 11.25.)
Based on examination of MagE spectra of spectrophotometric standard stars,
we chose not to use spectra from the bluest order of MagE, and 
not to use spectra $<3200$~\AA\ in the next order.
Due to detector fringing, the pipeline does not extract 
the two reddest orders of MagE.
Therefore, we extract spectra over the wavelength range 3200~\AA\ to 8280~\AA.  


The spectra of six \megasaura\ galaxies have coverage blueward of the Lyman limit 
at rest-frame 912~\AA . We have summed the flat-fielded, sky-subtracted, two-dimensional spectra for each
of these galaxies, and note the bluest wavelength where we see the continuum trace; 
in each case, we see flux down to the Lyman limit, but not blueward.
Therefore, we trim the final 1D spectrum of each galaxy at an observed wavelength
that is the greater of 3200~\AA\ or $(1+z)912$~\AA .

To measure the instrumental line spread function, which determines the spectral resolution, we
extract and combine the spectra of the night sky line emission in the same way as the object spectra, 
using the same weights applied to the object spectra.  The only difference is that when combining
the sky spectra, the observed wavelengths receive no barycentric correction.  From the final
sky spectra, we identify isolated, bright night sky emission lines, 
fit each with a Gaussian, and measure the median resolution over the spectrum, 
where the resolution $R \equiv \lambda / \delta\lambda$, with 
$\delta\lambda$  defined as the full width at half maximum.    
The spectral resolving powers range from R$=$2500 to R$=$4700, 
with an average of \avgR , and are tabulated in Table~\ref{tab:sample}. 
For combined spectra taken with the 2\arcsec\ slit, the median resolution 
and standard deviation are  $R=2545 \pm 37$; 
the observations taken with the 1\arcsec\ slit have $R=4770 \pm 8$ .
As expected for an echellete, the spectral resolution is constant with wavelength. 
These measured resolutions are higher than the design resolution of $R\sim4100$ for a 1\arcsec\ slit 
\citep{Marshall:2008bsa}.

We correct the flux density of each spectrum for Milky Way reddening, using the $E(B-V)$ value 
derived from Pan-STARRS~1 and 2MASS photometry by 
\citet{Green:2015cf},\footnote{Queried using the API provided by those authors at http://argonaut.skymaps.info}
assuming an extinction--to--reddening constant of 
$R_v = 3.1$, and the reddening curve of \citet{Cardelli:1989dp}.  
Table~\ref{tab:sample} lists the $E(B-V)$ values used.

As discussed above, spectra were only obtained for a portion of each lensed galaxy.  
The fluxing is therefore appropriate for the observed portion of the galaxy.  
To illustrate this, we use the example of \stwelve.  
\citet{Koester:2010ky} improved on the SDSS~DR7 photometry through
careful subtraction of neighbors, measuring  
($g$,$r$,$i$) AB magnitudes of 21.14, 20.60, and  20.51.
Using kcorrect \citep{Blanton:2007kl}, we measure the SDSS 
AB-magnitudes of the 
flux--calibrated MagE spectrum to be g,r,i $=$     22.15, 21.70, and  21.62.
The offset between these measurements implies that the slit captured 
$39\%$, $36\%$, and $36\%$ of the total light in $g$,$r$,$i$ from this lensed galaxy.  

\subsection{Continuum fitting}\label{sec:contfit}
Continuum fitting is necessarily a somewhat arbitrary process. At the high signal-to-noise and spectral
resolution of \megasaura, the problem is compounded because the so-called continuum is not 
in fact featureless.    Rather, it is the sum of the rest-frame ultraviolet spectra of many hot stars, 
and as such contains numerous weak photospheric absorption features.  Different continua are required
for different purposes; we have therefore computed several different kinds of continuum fits.

\begin{enumerate}
\item Hand-fit smooth continua:  
We interactively fit splines to the continuum of each galaxy, using the x\_continuum tool in the XIDL package.
The continuum was fit from spectral regions that are free from expected emission or
absorption lines at the galaxy redshift (including Lyman $\alpha$),  
intervening absorption lines, or contamination from night sky lines.  
x\_continuum also employs iterative sigma rejection to remove spectral features.
We exclude from the fit the spectral region within $\lambda_r \pm 30$~\AA\ of  C~IV 1548, 1551~\AA.
For each galaxy, we fit the continuum three times, looping over the sample to minimize human error; 
we then adopt as the continuum the mean of the three fits, and as the uncertainty the error in the mean.  
This process allows us to quantify the uncertainty due to the subjectivity of manual continuum fitting.  
The resulting continuum 
is smooth and has no knowledge of the underlying stellar emission; it is therefore most 
appropriate for analyzing the velocity profiles of transitions tracing the interstellar medium.

\item Automatic continua:  We automatically generate smooth continua as follows.  
For each spectrum, after masking out bad pixels as well as
regions with velocities within $500$~\kms\ of known emission and absorption features,
we smooth by convolving with a boxcar of width $\lambda_{rest}=100$~\AA.
The resulting continuum is smooth and has no knowledge of the underlying stellar emission.  
Since this continuum is automatically generated, it can be applied to a large number of 
theoretical models.

\item Stellar continua: We have generated best-fit
  spectra from linear combinations of Star- burst99 models 
\citep{Leitherer:1999jt, Leitherer:2010dz, Leitherer:2014ia}, 
each an instantaneous starburst with a  given age (between 1$-$40~Myr) and stellar continuum metallicity
  (with metallicities of 0.05, 0.2, 0.4, 1.0, and 2.0~Z$_\odot$) ,
  while simultaneously fitting for a stellar extinction ($E(B-V$)) from
  the line of sight dust reddening using a Calzetti extinction law \citep{Calzetti:2000iy}.
This fit is made assuming a linear  combination of the 50 stellar continuum models, and fitting for the
  linear coefficient (the fraction of the total light attributed to
  each Starburst99 model) using a least-squares method following the
  methodology of \citet{Chisholm:2015fo}.  Regions within 500~\kms\ of known emission
and absorption features were masked before fitting.
We use fully theoretically Starburst99 models that are computed using the Geneva stellar
  evolution tracks with high mass-loss \citep{Meynet:1994tx},  computed
  using the WM-BASIC method \citep{Leitherer:2010dz}, and assuming a
  Kroupa IMF with a high (low) mass power-law index of 2.3 (1.3) with
  a high-mass cut-off of 100~M$_\odot$. This modeling, and derived
  constraints on the stellar populations, will be described in a
  sub-sequent paper (Rigby et al.\ in prep.) that comprehensively
  analyzes the diagnostics of hot stars in the \megasaura\ spectra. What
  is relevant for the current paper is that these models fit the underlying
  stellar spectra, and can therefore be used to estimate the redshift
  of the stars, commonly known as the systemic redshift.

\end{enumerate}

\subsection{Measurement of redshifts}\label{sec:redshifts}
We measure the redshifts of the stars, the nebulae, and the interstellar medium for each galaxy, 
as follows.

The stellar redshift is an output from the Starburst99 fitting (Rigby \etal\ in prep.), 
in which we cross-correlate the model and the observed data to find the most likely stellar
redshift.  We shift the Starburst99 models in velocity-space by 10~km~s$^{-1}$ increments 
over a range of $\pm600$ km~s$^{-1}$ from the measured nebular redshift. 
We then produce a likelihood function by taking the inverse exponential of the 
$\chi^2$ function using the observed data and the shifted Starburst99 models. 
We calculate the likelihood at each velocity shift, normalize the likelihood
function, and take the expectation value and standard deviation of
this probability density function as the best-fit redshift and the redshift uncertainty.  
As such, the quoted redshift uncertainty does not take into account the 
full range of possible stellar models.

For 12 of the spectra in the \megasaura\ sample, the nebular redshift is 
measured from the [C~III]~1907,  C~III] 1909~\AA\ doublet, 
since these are generally the brightest rest-frame UV nebular lines.   
We determine a redshift for the  [C~III]~1907,  C~III] 1909 feature by fitting two Gaussians with a 
common linewidth, holding fixed the ratio of the central wavelengths, using the IDL 
Levenberg–Marquardt least-squares fitting code MPFITFUN \citep{Markwardt:2009wq}, 
following  \citet{Rigby:2014hq}.  The uncertainty quoted is the uncertainty from MPFITFUN.

Table~\ref{tab:redshifts} notes that for  5 spectra,  the nebular redshift is
sourced differently, from nebular lines in the rest-frame optical.  While lines like
H$\beta$ and [O III]~5007~\AA\ generally have higher fluxes than  [C~III]~1907,  C~III] 1909\AA,
using them may introduce systematic wavelength calibration errors when comparing to 
stellar redshifts measured from the rest-frame UV continuum.

To test for systematic effects, we measured redshifts individually for [C~III], C~III], 
and weaker rest-frame UV emission lines 
( O~III]~1666, Si~III]~1882, Si~III]~1892, C~II]~2325 , and [O~II]~2470), 
in four spectra with high signal-to-noise ratio 
(RCS0327-knotE,  S0004-0103, S0108+0624, and S0957+0509.) 
We note a systematic effect, that compared to other rest-frame UV emission lines,  
[C III]~1907 is blueshifted in all cases by up to 40~\kms, while
in three of four cases C III]~1909 is redshifted by up to 20~\kms. 
The standard deviation is 10--20~\kms.
This effect appears to be real; we do not understand its origin.  
Perhaps the forbidden [C III] 1907 and semi-forbidden C~III] 1909 arise in 
different physical regions with different velocities.  Perhaps 
the line profiles are intrinsically asymmetric, driven by winds, and 
the bias comes from fitting them with Gaussians.  
In any case, we mitigate this bias by using both transitions to determine the redshift,
as described above.
 
The interstellar medium redshift is measured as the average redshift of 
Gaussian fits to C~II~1334, Si~II~1526, Al~II~1670, Al~III~1854 and Al~III~1862; 
the uncertainty is the standard deviation.
The measured redshifts are tabulated in Table~\ref{tab:redshifts}.

\section{Results}\label{sec:results}

\subsection{The \megasaura\ spectra}\label{sec:prettyspectra}
Figure~\ref{fig:snr} shows the per-pixel signal-to-noise ratios (SNR) for the final combined \megasaura\ 
spectra.  The median per-pixel dispersion is 0.35~\AA\ in the observed frame.  
A total of 8 spectra have per-pixel $SNR>10$ over the observed
wavelength range of 5000--7000~\AA ; the median spectrum has a per-pixel $SNR \sim 10$.
For the average resolution, at 5000~\AA\ this corresponds to 
$SNR=21$ per resolution element.

The final combined \megasaura\ spectra are plotted in Figure~\ref{fig:spectra}. 
The \megasaura\ spectra comprise a spectral atlas at ``cosmic noon'' that has 
superior signal-to-noise ratio and wavelength coverage, albeit lower spectral resolution, 
than the COS/\textit{HST} archive of starburst galaxies in the local universe.

\subsection{Contribution from Active Galactic Nuclei}\label{sec:AGN}
Two galaxies in our sample show evidence of active galactic nuclei (AGN).
The MagE spectrum of SGAS~J224324.2$-$093508 shows broad emission lines of Lyman~$\alpha$, 
N~V, C~IV, and [C~III], that together indicate that this galaxy hosts a broad-line AGN.
The \textit{HST} imagery  (Figure~\ref{fig:where_pointed}) shows 
a spatially extended host galaxy with a central point source.
The MagE spectrum of SGAS~J003341.5$+$024217 does not show obvious AGN diagnostics.
However, a VLT/SINFONI observation shows that the nucleus exhibits high-ionization 
[O III]~5007 / H$\beta$ and H$\alpha$/[N~II] ratios, and also broad wings in H$\alpha$ 
(E.~Wuyts, private communication.)  It is thus implicated as having a weak AGN.

\subsection{Velocity offsets}\label{sec:velresults}
Figure~\ref{fig:wind_offsets} compares the velocity offsets among the 
interstellar medium, stars, and nebulae of the \megasaura\ spectra, 
measured as described in \S~\ref{sec:redshifts}, and tabulated in Table~\ref{tab:redshifts}.
The top left panel of Figure~\ref{fig:wind_offsets} shows that there is 
no systematic offset between the redshifts of the hot stars and the nebulae; the median offset 
and median absolute deviation are $-1 \pm 31$~\kms ;  
from bootstrapping, the $80\%$ confidence interval on the median is $-1^{+20}_{-11}$~\kms .

The top right panel of  Figure~\ref{fig:wind_offsets} shows that the 
ISM lines are systematically blue-shifted with respect to the stellar continuum, 
with a median offset and median absolute deviation of
$-148 \pm 35$~\kms,  and 
bootstrapped $80\%$ confidence interval on the median of $-148^{+42}_{-11}$~\kms .
This is to be expected given the ubiquity of galaxy-wide winds in z$\sim$2 galaxies 
\citep{Shapley:2003gd, Weiner:2009cf, Rubin:2014hv}.


We now place these measurements in a larger context. 
Because purely stellar features like the photospheric absorption 
lines are only detectable with high signal-to-noise and moderate spectral resolution, 
it is common practice in the literature to use nebular emission lines as a proxy for 
the systemic redshift.  Doing so makes two assumptions---that the ionized gas is well-mixed
with the hot stars, and that outflows do not seriously bias the nebular line redshifts.
The local galaxy NGC~7552 shows that such biases can occur:  
H$\alpha$ is blueshifted from the stellar redshift by  $-30$\kms\ due to a wind,
but this blueshift is much less than the bulk outflow velocity as traced
by rest-frame UV absorption lines.
These assumptions are testable for the \megasaura\ spectra.
In \S\ref{sec:redshifts} we reported that the brightest emission lines in the rest-frame 
UV, [C III]~1907 and C~III]~1909~\AA, may be systematically biased by up to
40~\kms\ with respect to each other, and with respect to still weaker rest-frame UV lines.  
Since one line is systematically blue-shifted and the other red-shifted, we attempted to 
mitigate the bias by jointly fitting the [C~III], C~III] doublet.  
Indeed, the median offset between the nebular redshifts and systemic redshifts,  $-1^{+20}_{-11}$~\kms \ 
($80\%$ confidence) is fully consistent with no offset.
This is consistent with the measurement of \citet{Shapley:2003gd} 
for their stack of Lyman Break Galaxies:  $-10 \pm 35$~\kms . 

We conclude that while nebular redshifts can be used as proxies for 
systemic redshifts, one should be aware of systematic biases at the 
level of tens of \kms\ when using only one line of the [C~III], C~III] doublet.
In the era of the \textit{James Webb Space Telescope} (\textit{JWST}), 
such biases may limit kinematic studies. 
The stellar redshifts we measure, by fitting the whole stellar continuum, 
should be unbiased to outflows, but are less precise than the nebular redshifts
(median fractional uncertainty of $3\times 10^{-4}$, versus $4\times 10^{-5}$).  
Greater precision can be obtained by fitting individual photospheric absorption lines, 
when they are clearly detected.  This is the case for about half the \megasaura\ spectra, 
and will be explored in a future paper.

\section{Summary}\label{sec:discussion}
In this Paper, we introduce Project \megasaura : \megasauralong.  
This is the sample and data release paper for that project, 
in which we present high-quality, medium-spectral resolution rest-frame ultraviolet 
spectra for \samplesize\ lensed galaxies.  
Given the wealth of spectral diagnostics in the  $1000< \lambda_r < 3000$~\AA\ spectral region
covered by these spectra, the \megasaura\ spectra should 
enable a number of investigations into the nature of star-forming galaxies at cosmic noon. 

Using the measured redshifts for the stars, the nebulae, and the interstellar medium 
in these galaxies, we analyze the relative velocity offsets.
Relative to the stars, the ISM gas shows an expected bulk outflow of $-148 \pm 35$~\kms\ 
(median and median absolute deviation).  
We see no offset between the stars and the nebular emission lines:   $-1 \pm 31$~\kms\
(median and median absolute deviation).  
We plan a future paper to analyze the multiphase outflowing winds and the
ionized nebulae at $z \sim 2$ as probed by the \megasaura\ spectra, and compare to 
galaxies in the local universe.

We have obtained the high-quality \megasaura\ spectra with an eye toward the future. 
The $1000< \lambda_r < 3000$~\AA\ spectral region is unusually rich in spectral diagnostics. 
This spectral region is what 20--30~m telescopes now under construction will probe 
for unlensed field galaxies at $z\sim2$; it is also the spectral region that the 
James Webb Space Telescope will probe for the galaxies in the epoch of reionization. 
Given this future, and as a service to the community, we publish 
electronic versions of the \megasaura\ spectra.  

\acknowledgments  Acknowledgments: 
This paper includes data gathered with the 6.5 meter Magellan Telescopes located at 
Las Campanas Observatory, Chile.
We thank the staff of Las Campanas for their dedicated service, which has 
made possible these observations.  We thank the telescope allocation committees of the Carnegie 
Observatories, The University of Chicago, The University of Michigan, and Harvard University, 
for supporting this observing program over several years.
This paper includes data from observations made with the Nordic Optical Telescope, operated 
by the Nordic Optical Telescope Scientific Association at the Observatorio del Roque 
de los Muchachos, La Palma, Spain, of the Instituto de Astrofisica de Canarias.

\bibliographystyle{astroads}
\bibliography{papers}  

\begin{figure*}[h!]
\includegraphics[height=2.2in]{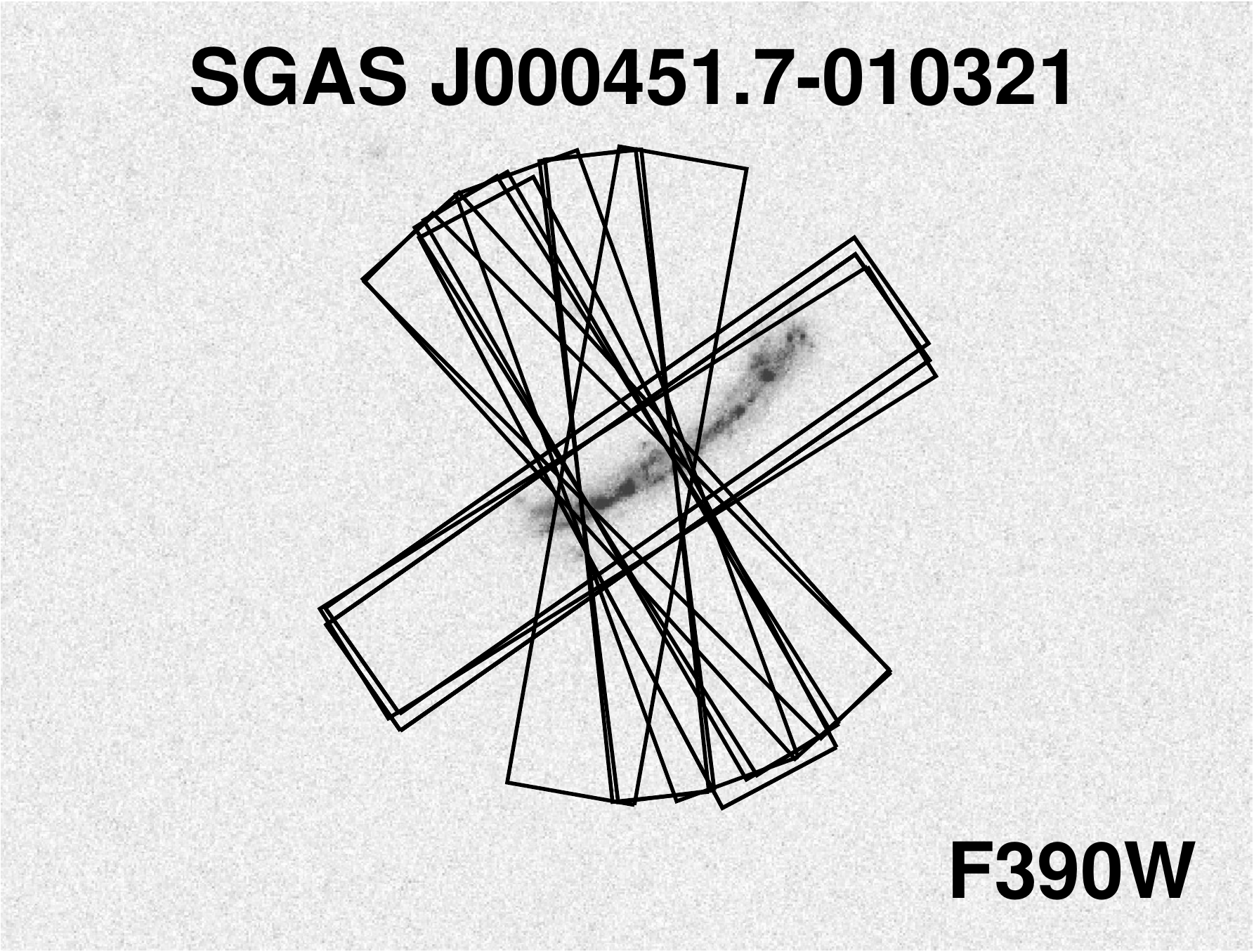} 
\includegraphics[height=2.2in]{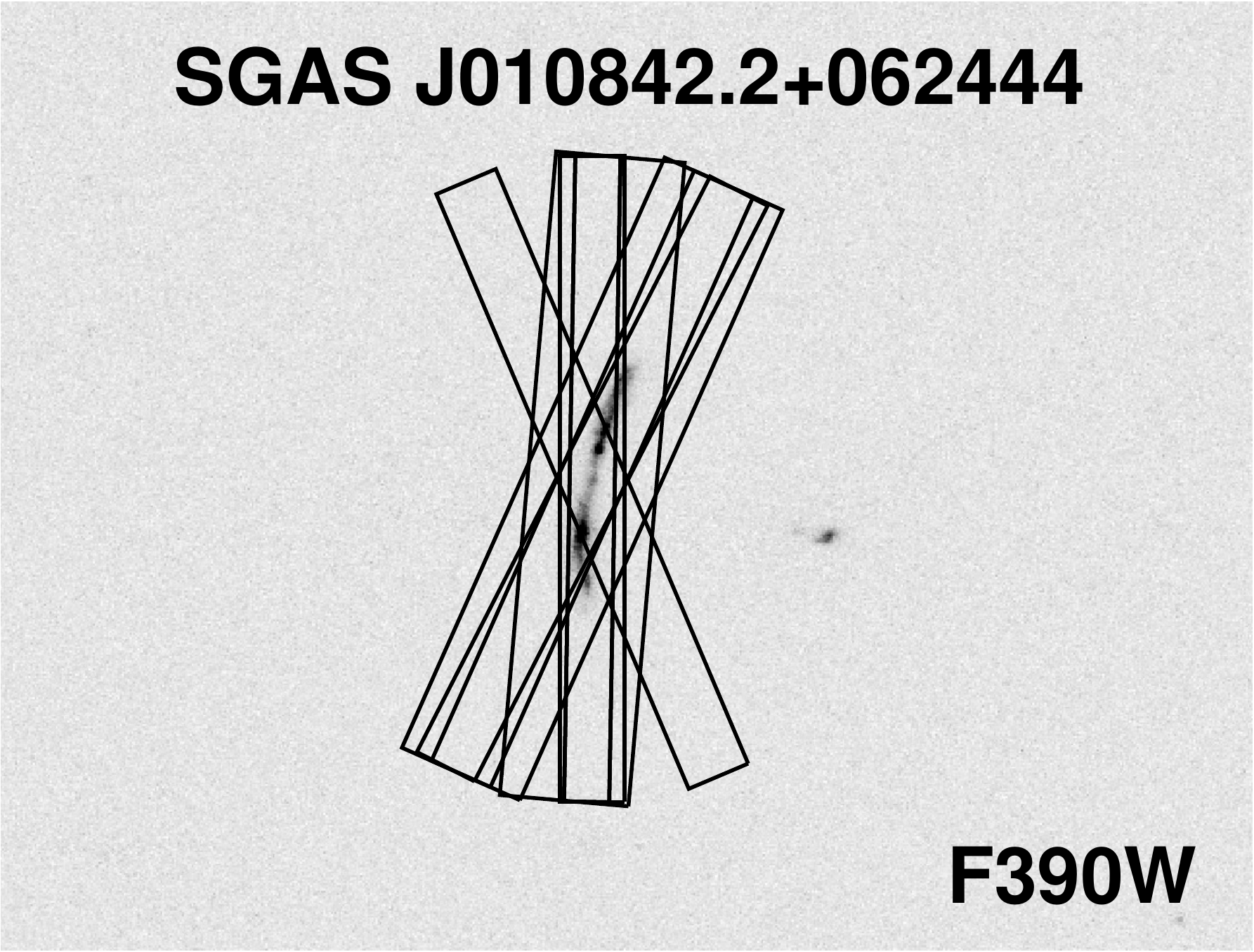} 

\includegraphics[height=2.2in]{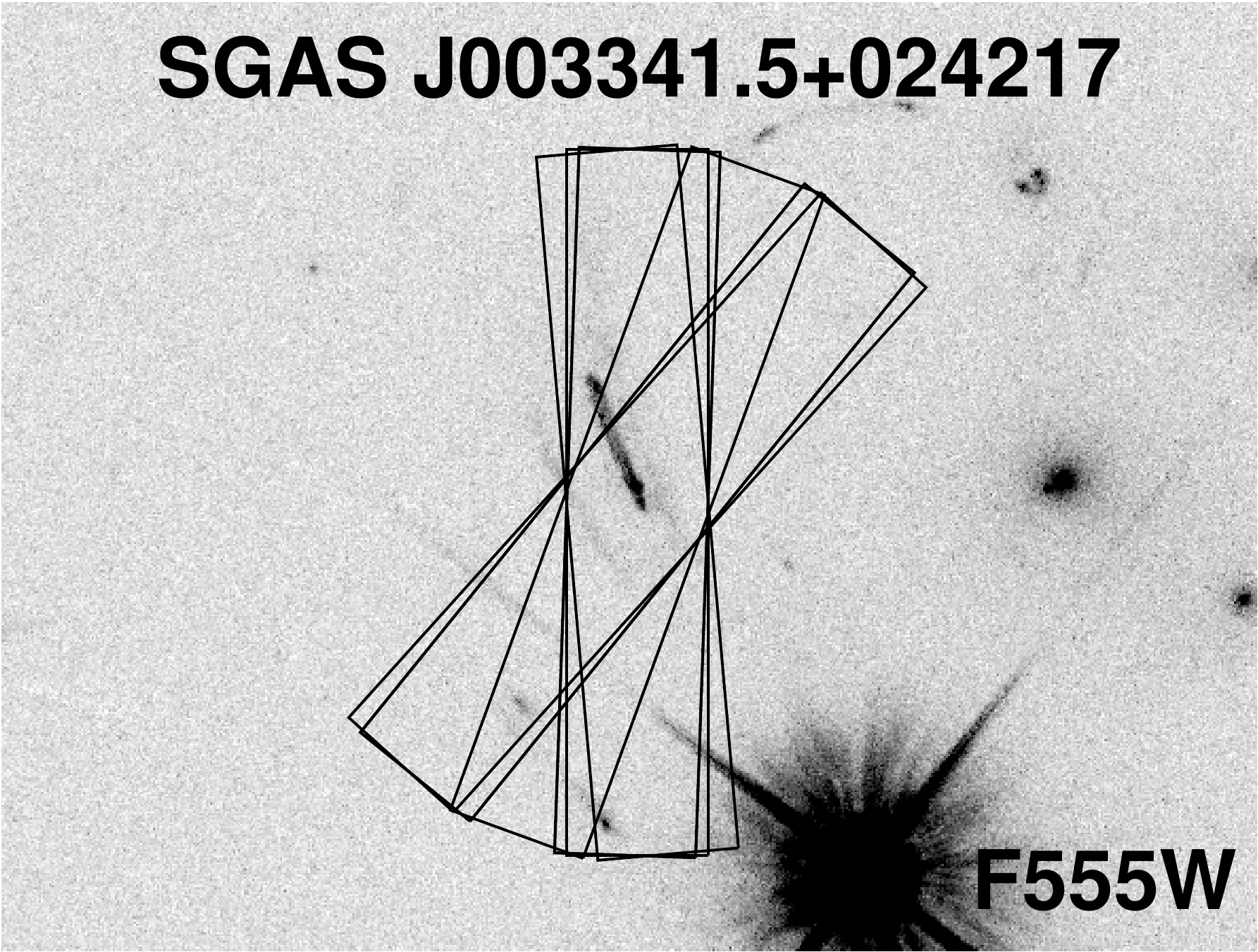} 
\includegraphics[height=2.2in]{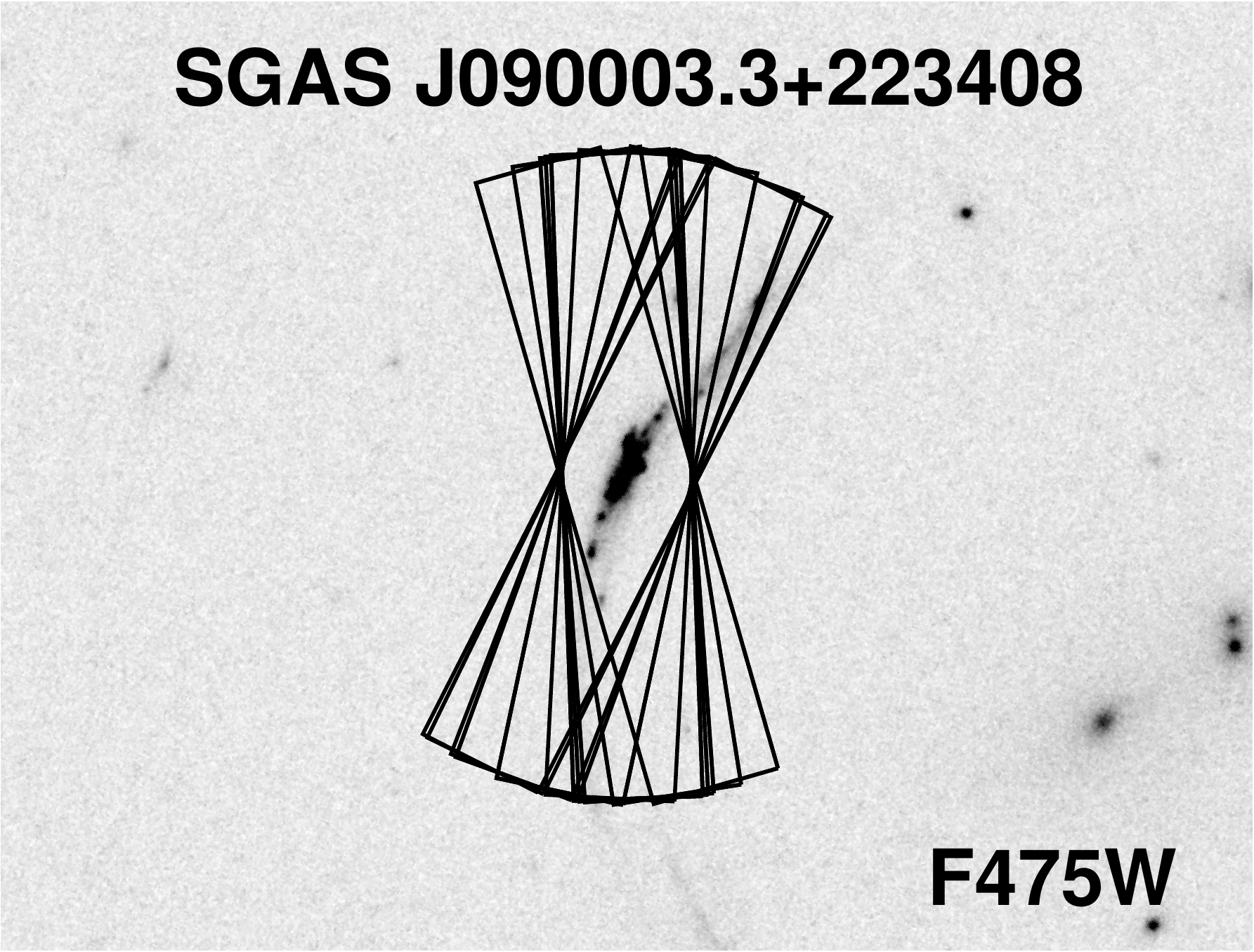} 

\includegraphics[height=2.2in]{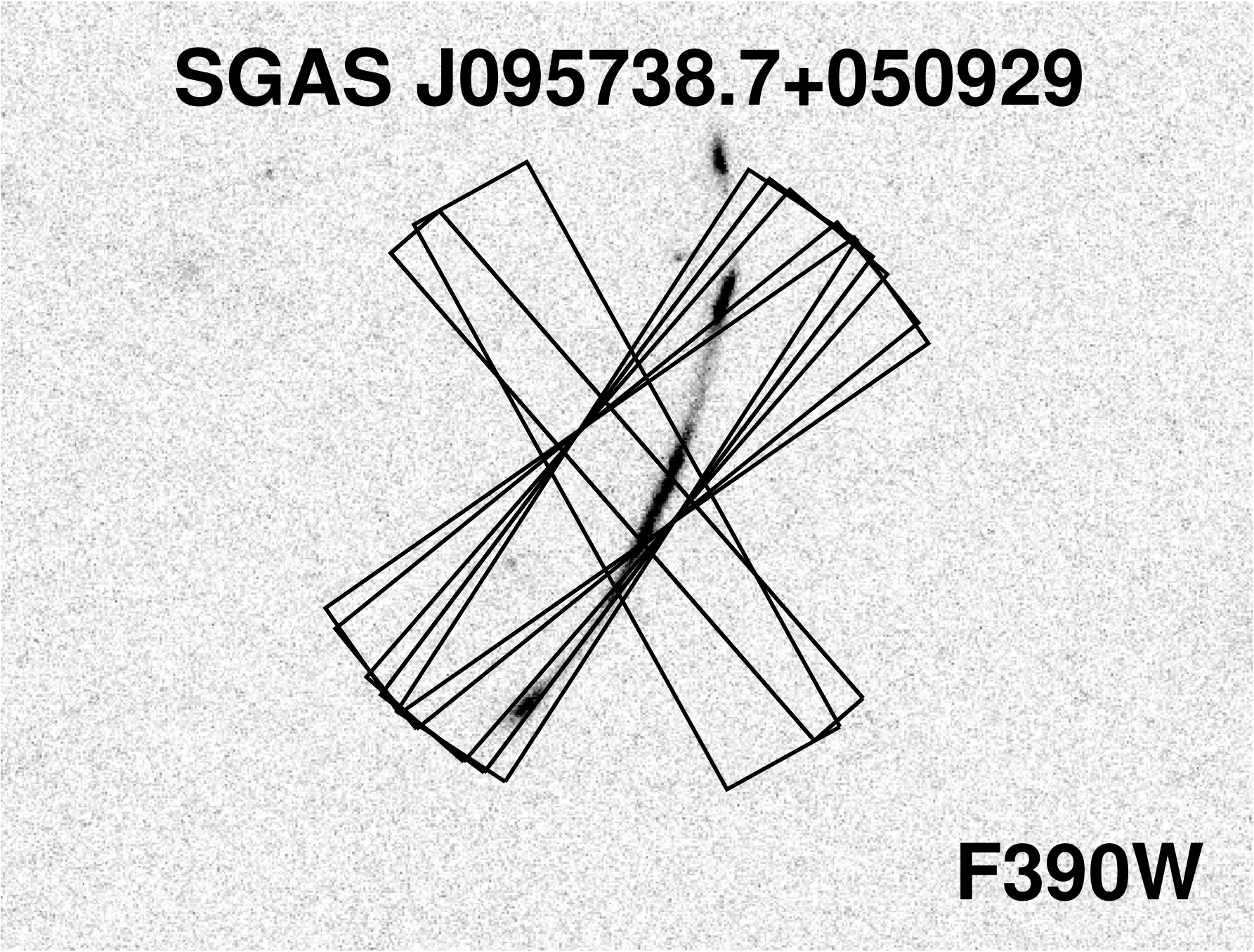} 
\includegraphics[height=2.2in]{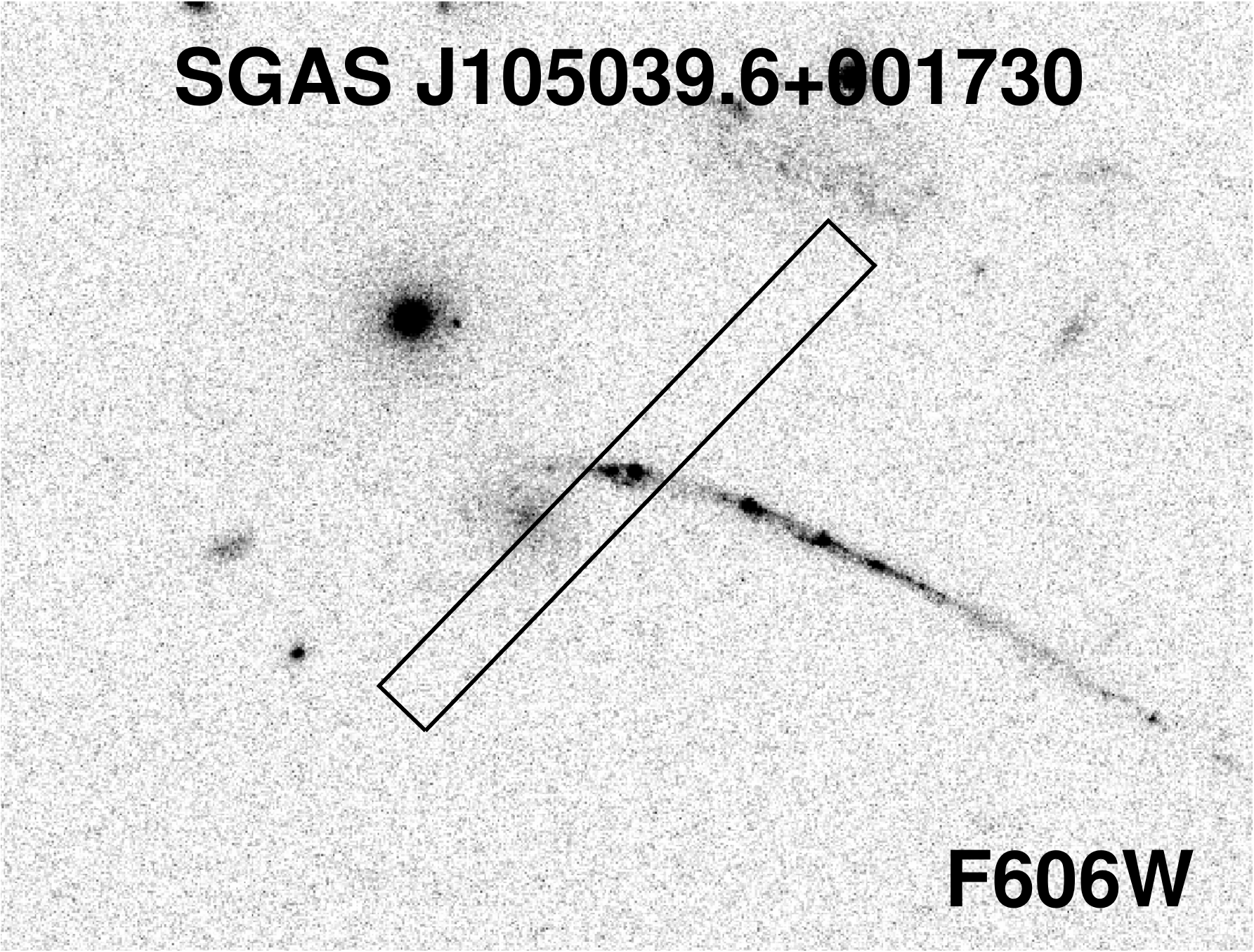} 

\includegraphics[height=2.2in]{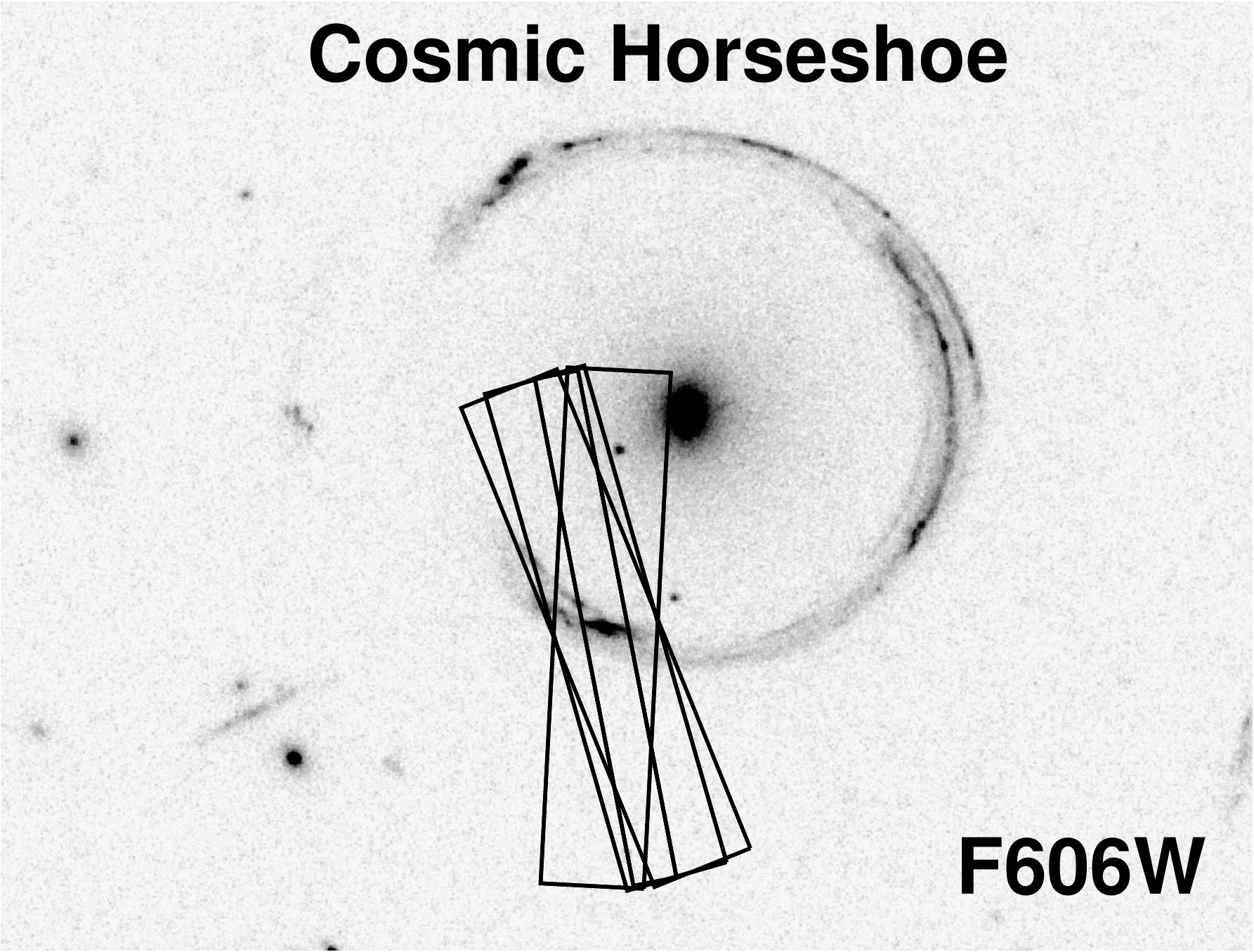} 
\includegraphics[height=2.2in]{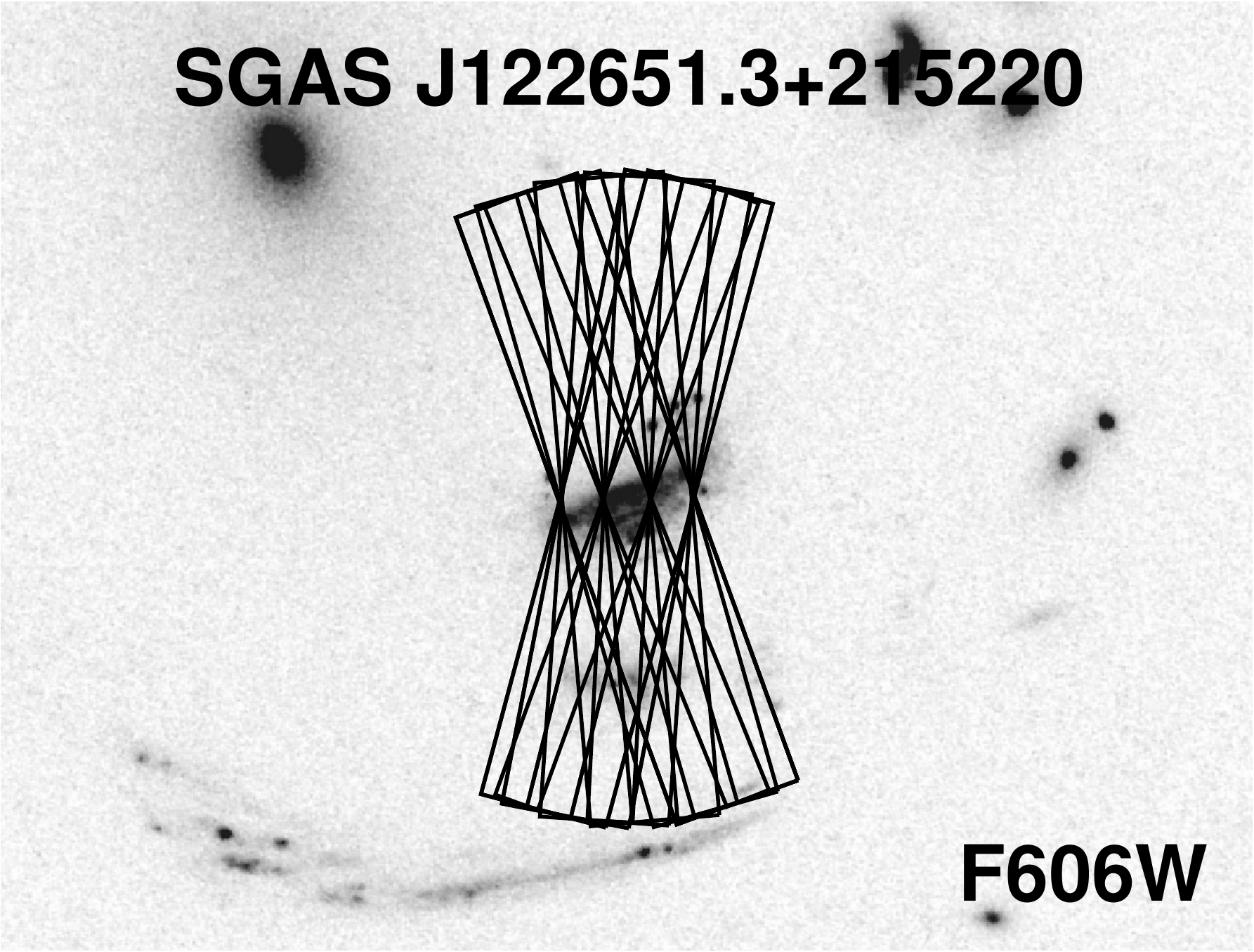} 
\end{figure*}

\begin{figure*}
\includegraphics[height=2.2in]{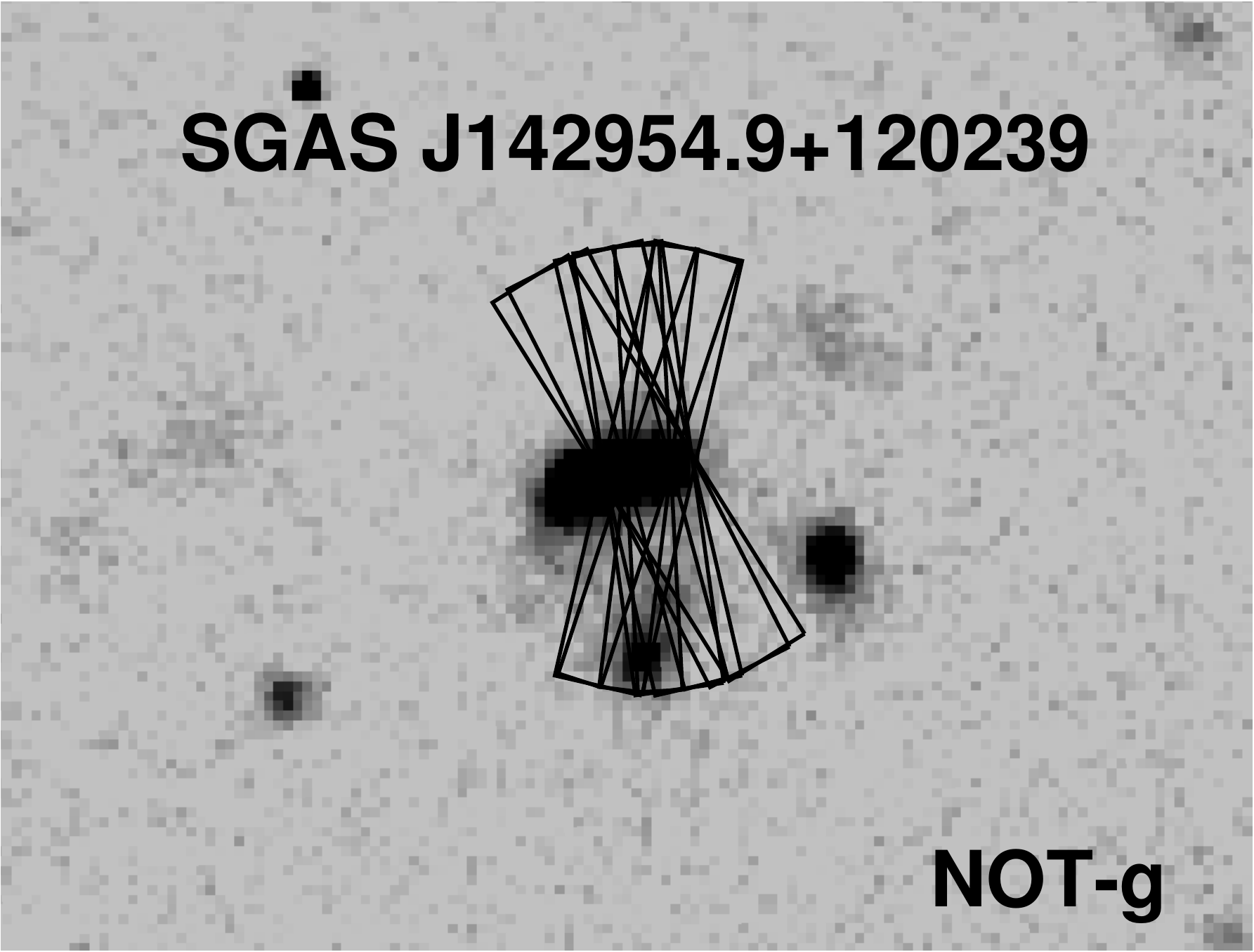} 
\includegraphics[height=2.2in]{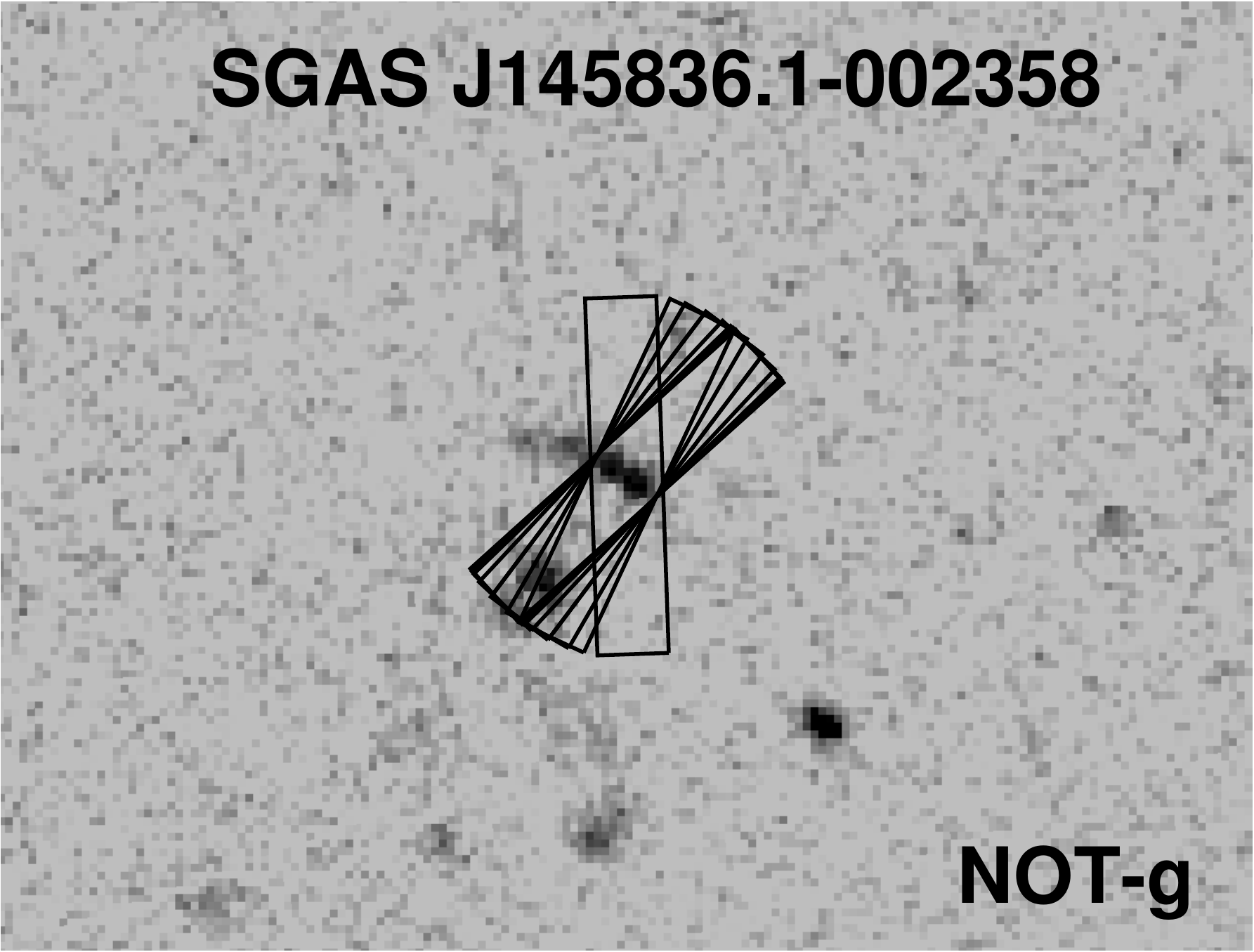} 

\includegraphics[height=2.2in]{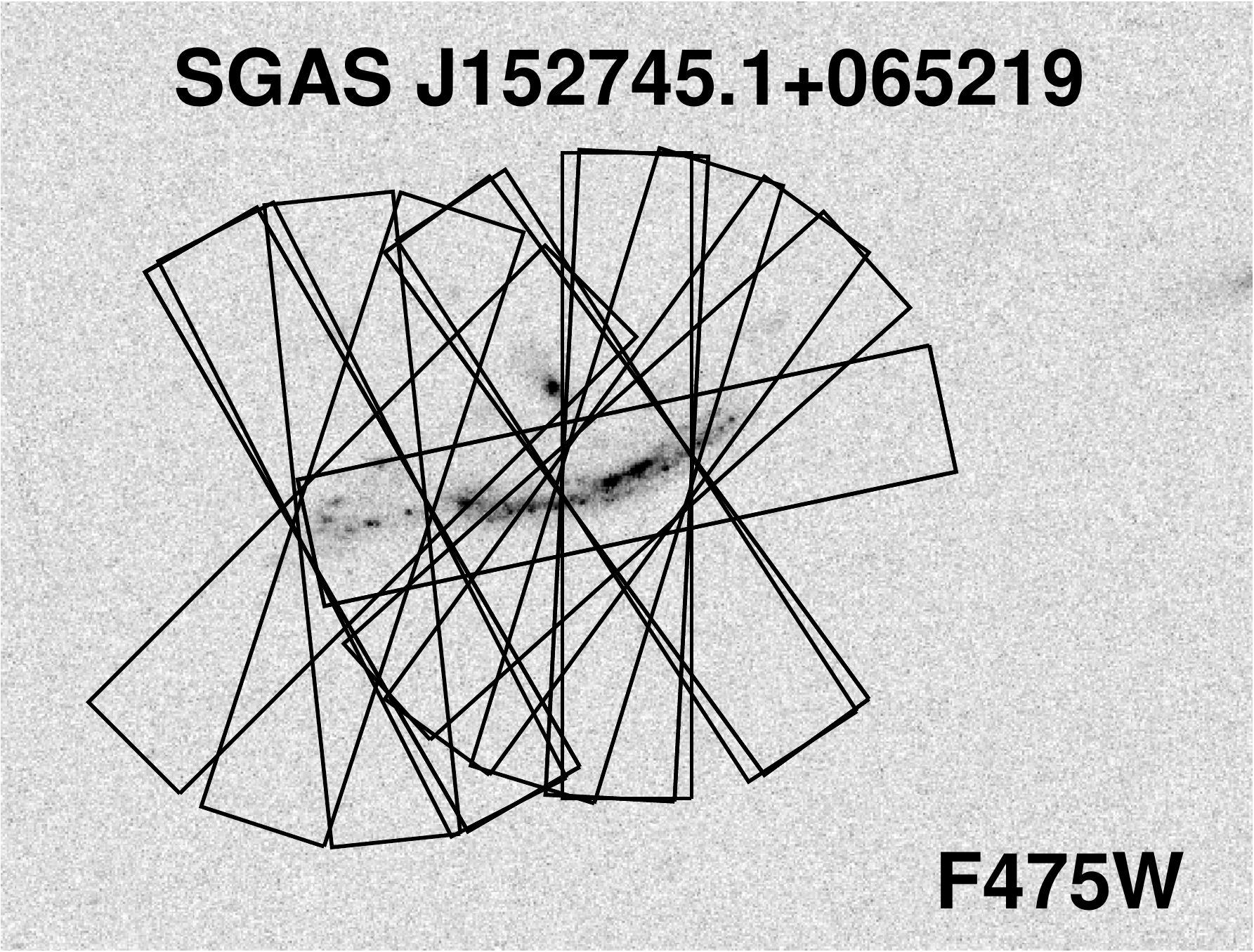} 
\includegraphics[height=2.2in]{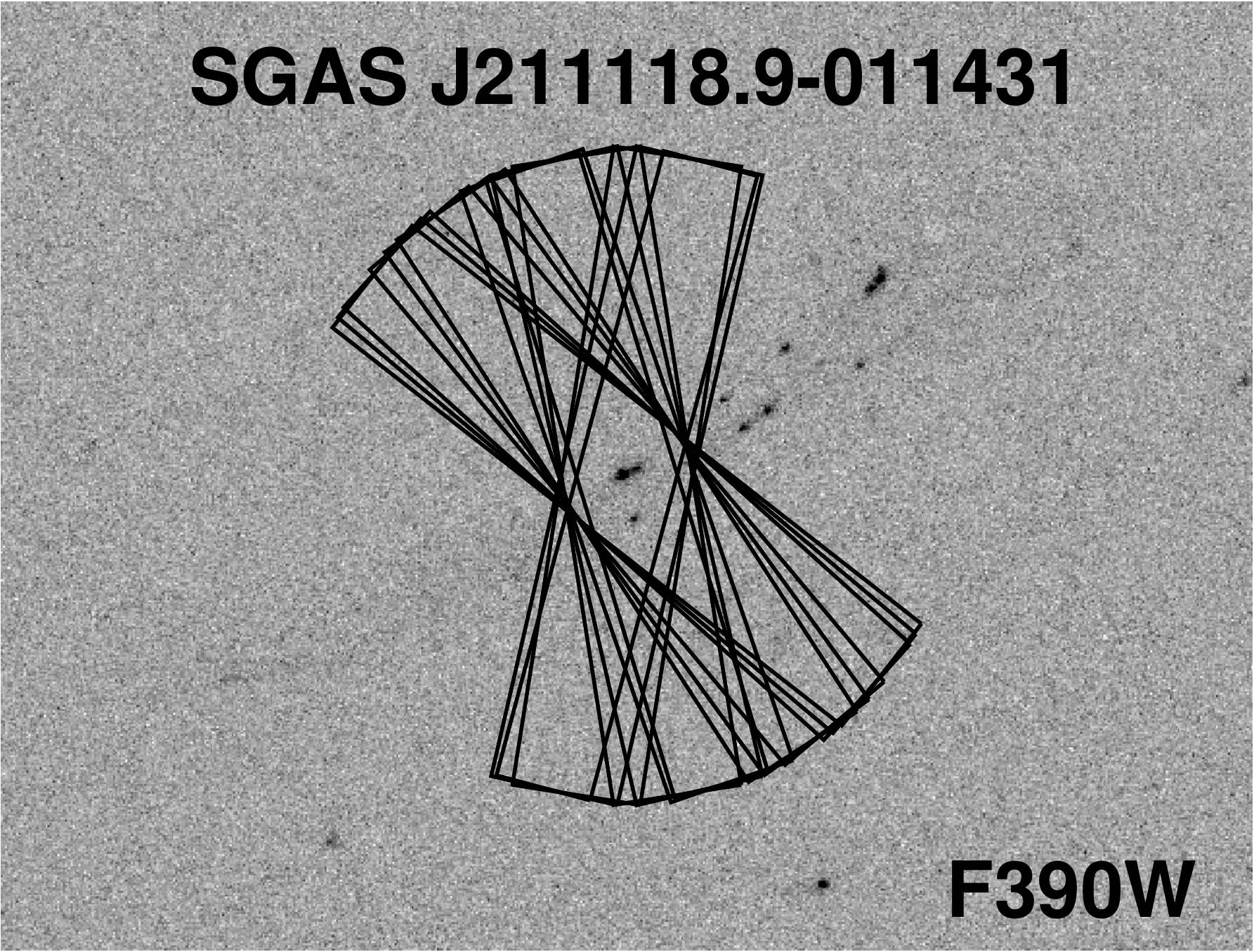} 

\includegraphics[height=2.2in]{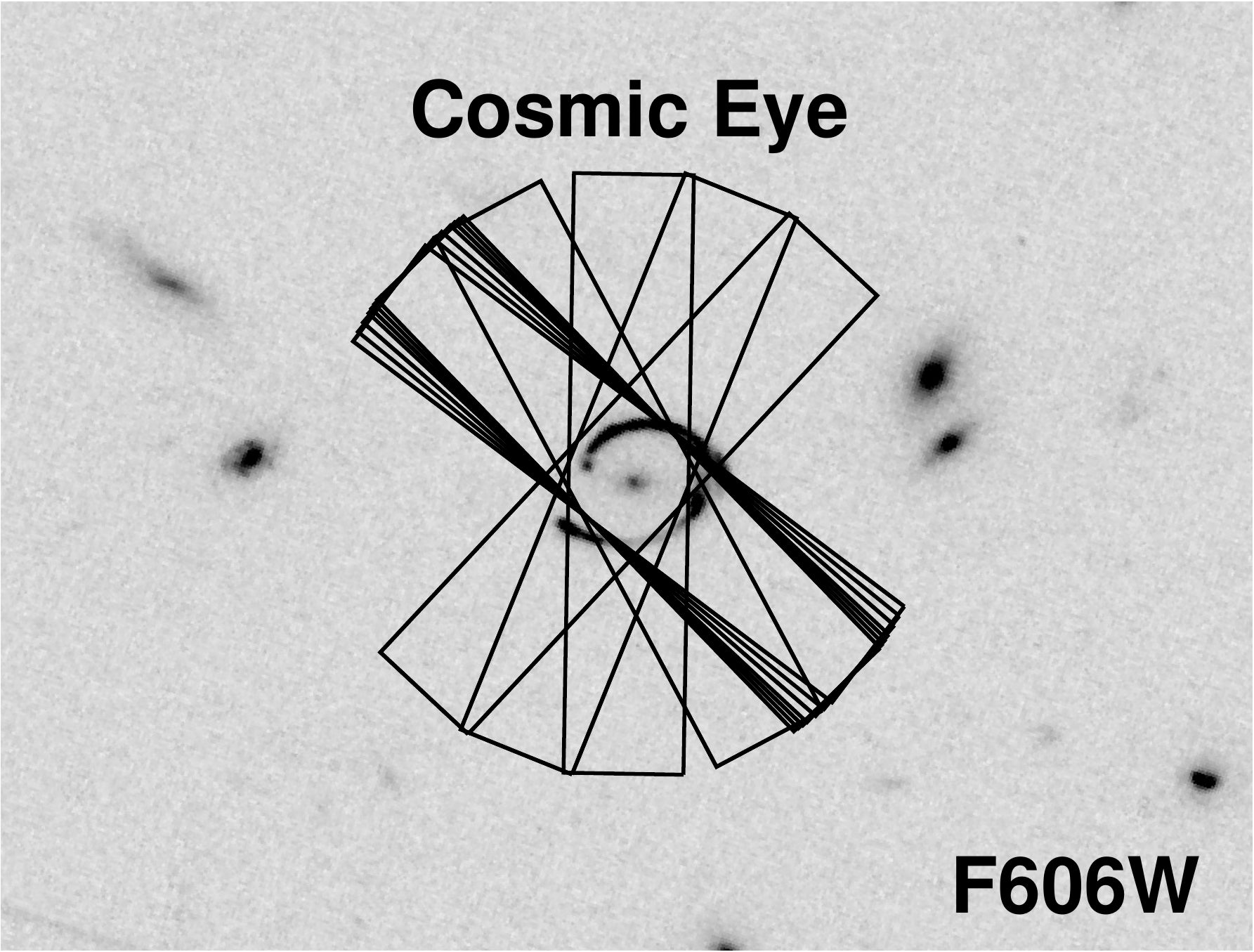} 
\includegraphics[height=2.2in]{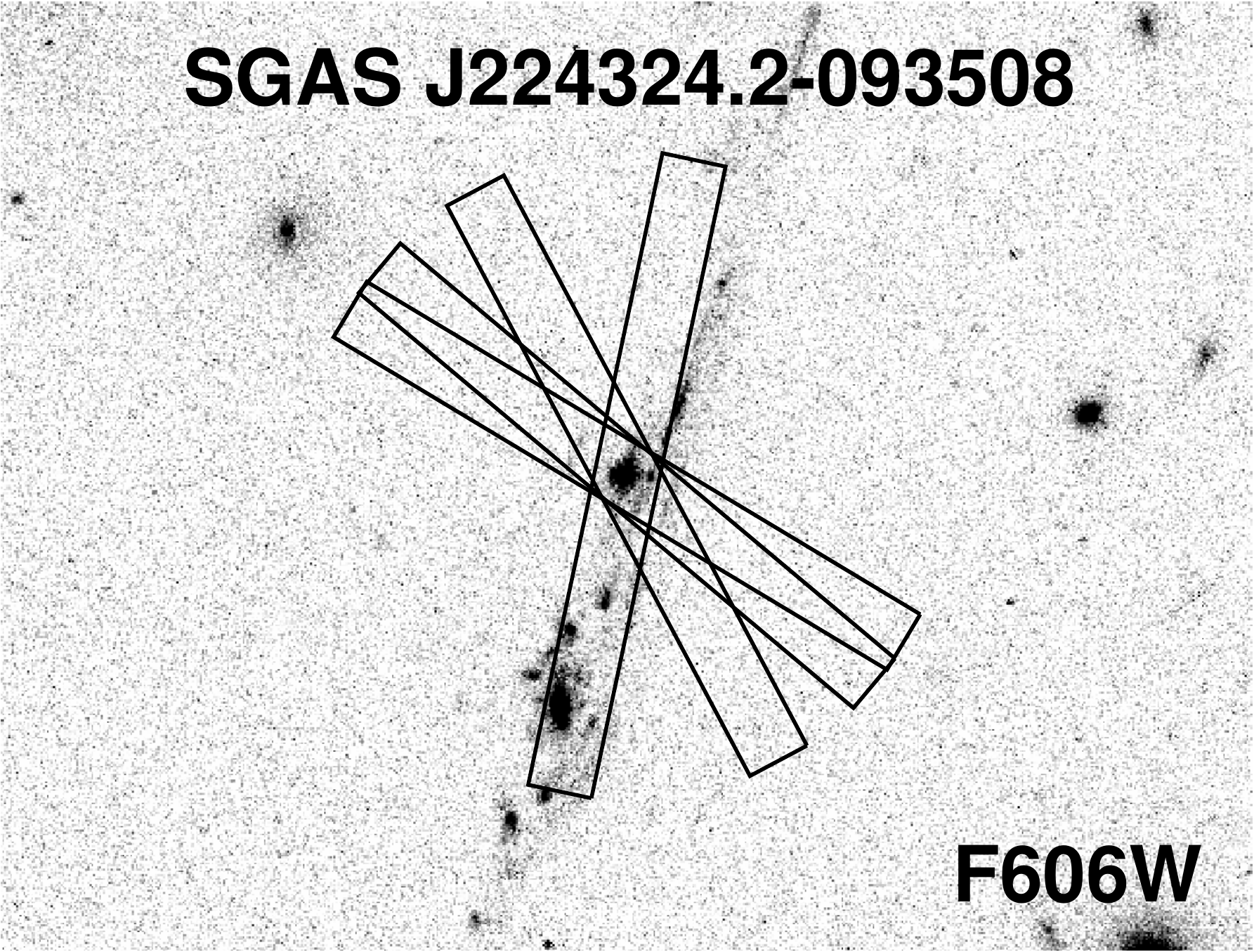} 
\figcaption{Findercharts indicate what portion of each lensed galaxy was targeted for MagE spectroscopy.  
\rcsohthree\ is plotted separately in Figure~\ref{fig:where_pointed_rc0327}.  
The filter of the background image used is indicated by text at the lower right of each image: 
F---W indicates an \textit{HST} image, generally the WFC3 instrument; 
NOT-g indicates a g-band image from  the Nordic Optical Telescope; 
A slit is drawn for each integration, using our best estimate of where that observation was pointed, 
as well as the actual slit position angle 
and slit width.  The length of the MagE slit is 10\arcsec .  For all images, N is up and E is left.
\label{fig:where_pointed}}
\end{figure*}

\begin{figure*}
\includegraphics[width=5.5in]{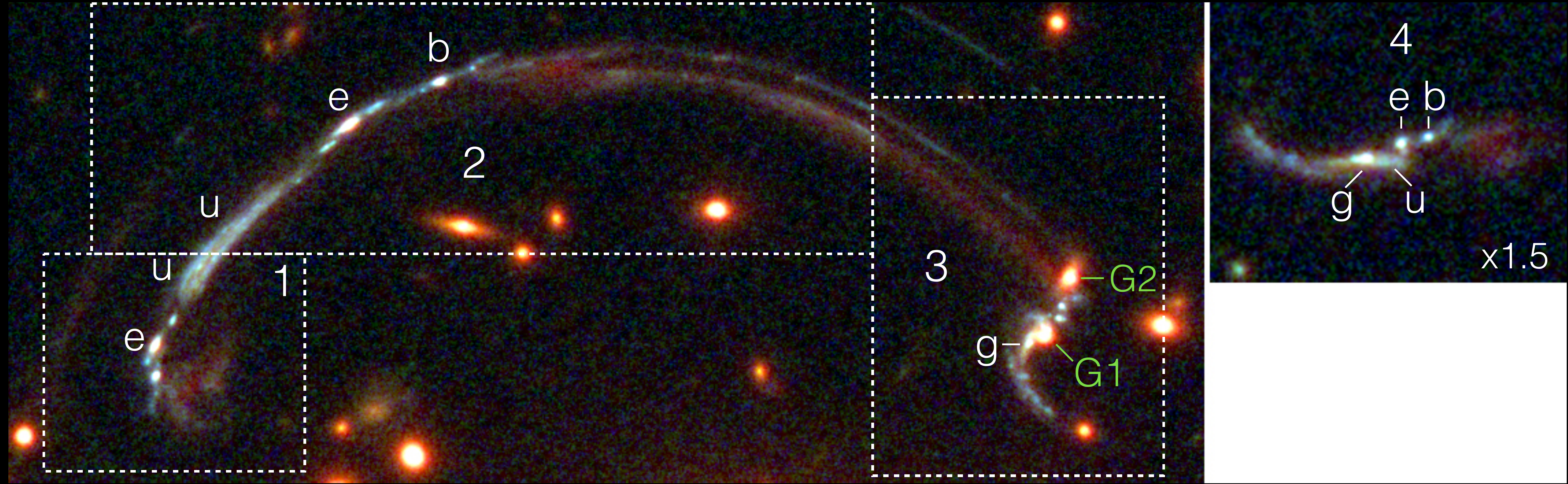} 
\includegraphics[width=5.5in]{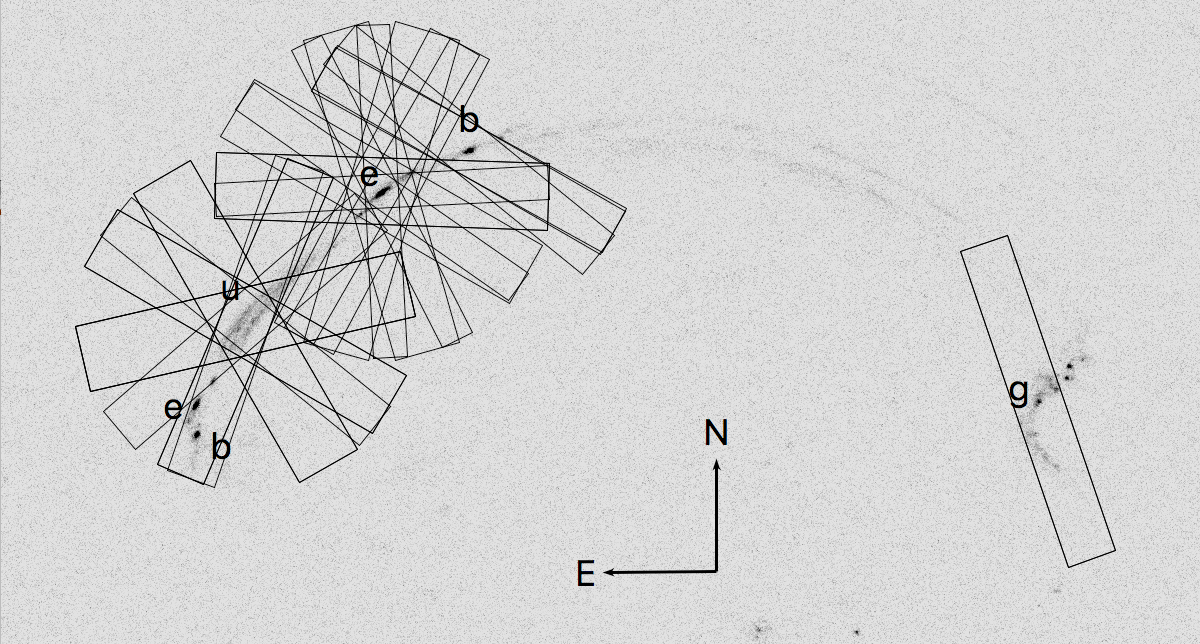} 
\includegraphics[width=5.5in]{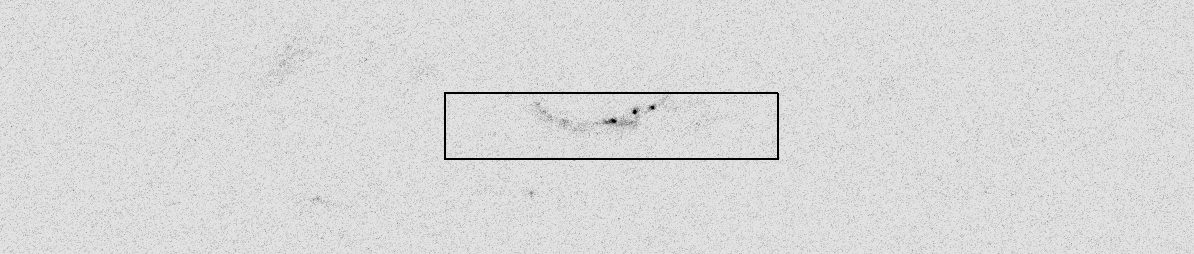} 
\figcaption{ Finderchart for the multiply imaged lensed galaxy RCSGA 032727-132609.  
TOP PANEL: Color rendition composed of
HST/WFC3   F160W, F125W, F098M (red); 
F814W, F606W (green); and F390W (blue).  The dashed lines approximately
mark the multiple images, indicated by numbers.  The emission knots 
with MagE spectroscopy are labeled as ``e", ``u", ``b", and ``g", following
the labeling of \citet{Sharon:2012dr} .
Two cluster galaxies ``G1" and ``G2" interrupt image 3.
The right panel shows a zoomed-in (by a factor of $\times 1.5$) 
view of the counter-image, which is a relatively undistorted image of the source-plane galaxy.  
Figure adapted from  \citet{Sharon:2012dr} .
MIDDLE AND BOTTOM PANELS: HST/WFC3 F390W image, with MagE observations overplotted.  One
slit is drawn for each integration, using our best estimate 
of the pointing, the actual slit position angle, and the
slit width used during that observation.  
Since the slit was regularly rotated to track the parallactic angle, 
knots U and E were observed using a wide range of slit position angles.  
The length of the MagE slit is a fixed 10\arcsec.   
North is up and East is left.
\label{fig:where_pointed_rc0327}}
\end{figure*}

\begin{figure*}
\includegraphics[page=1,scale=0.7]{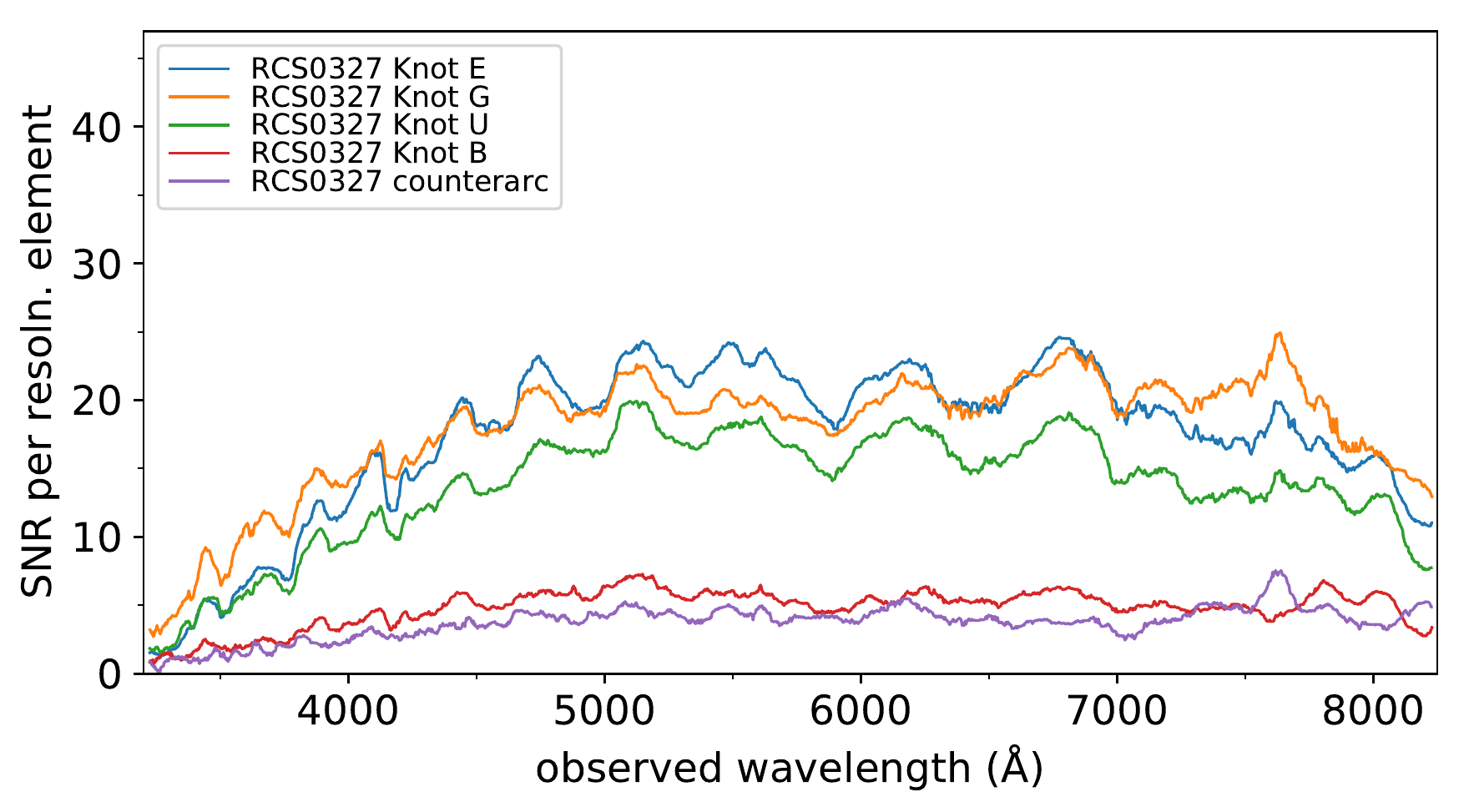} 

\includegraphics[page=2,scale=0.7]{f3.pdf} 

\includegraphics[page=3,scale=0.7]{f3.pdf} 
\end{figure*}

\begin{figure*}
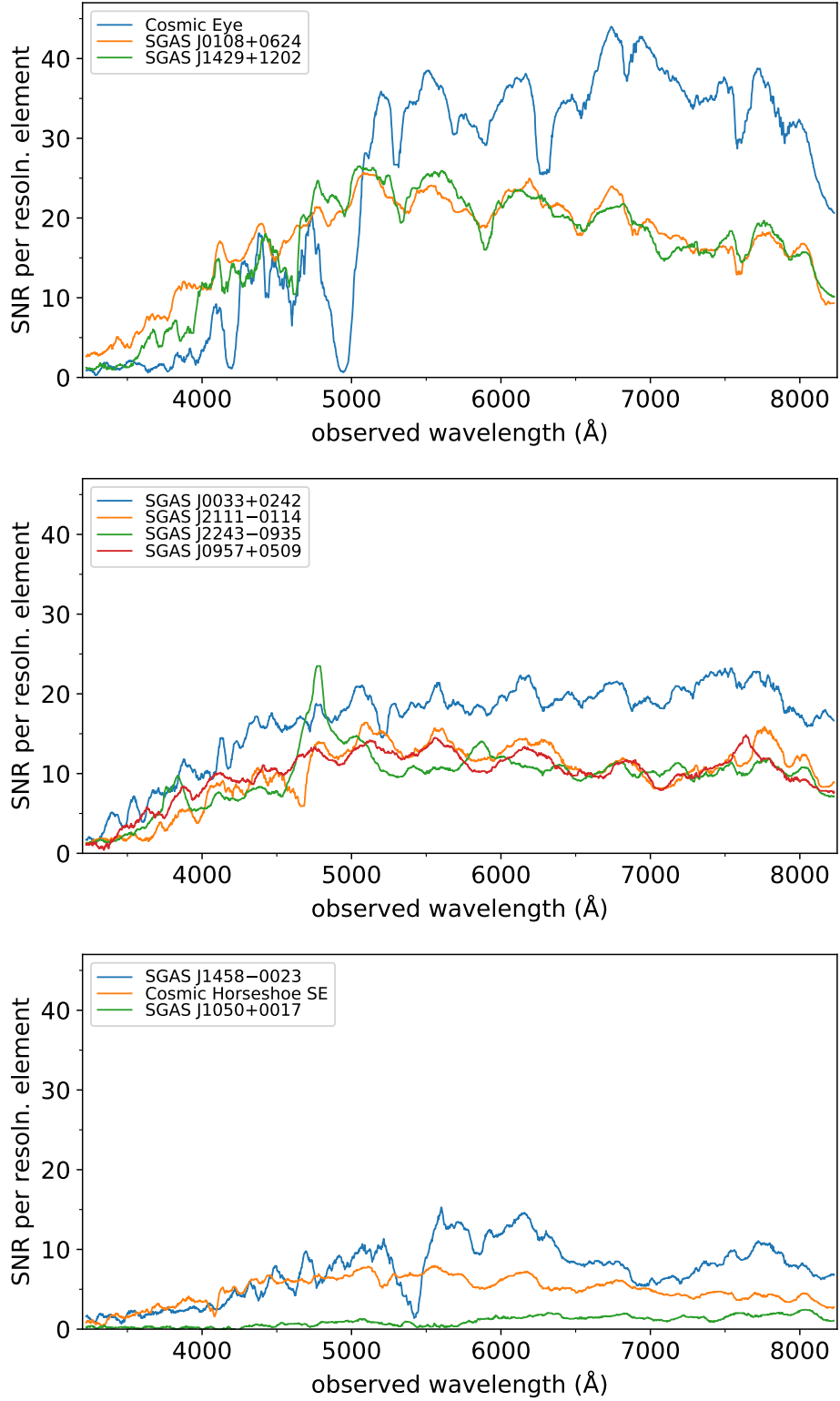

\includegraphics[page=4,scale=0.7]{f3.pdf} 

\includegraphics[page=5,scale=0.7]{f3.pdf} 

\includegraphics[page=6,scale=0.7]{f3.pdf} 
\figcaption{Signal-to-noise ratio of the spectra.  
For each spectrum, we plot the
signal-to-noise per resolution element, smoothed for readibility, 
versus observed wavelength.
We plot in separate panels the two galaxies, \rcsohthree\ and \sfifteen, 
that have spectra of multiple physical regions.  
The other galaxies are plotted in order of decreasing signal-to-noise ratio.
\label{fig:snr}}
\end{figure*}

\begin{figure*}
\includegraphics[page=1,scale=0.4]{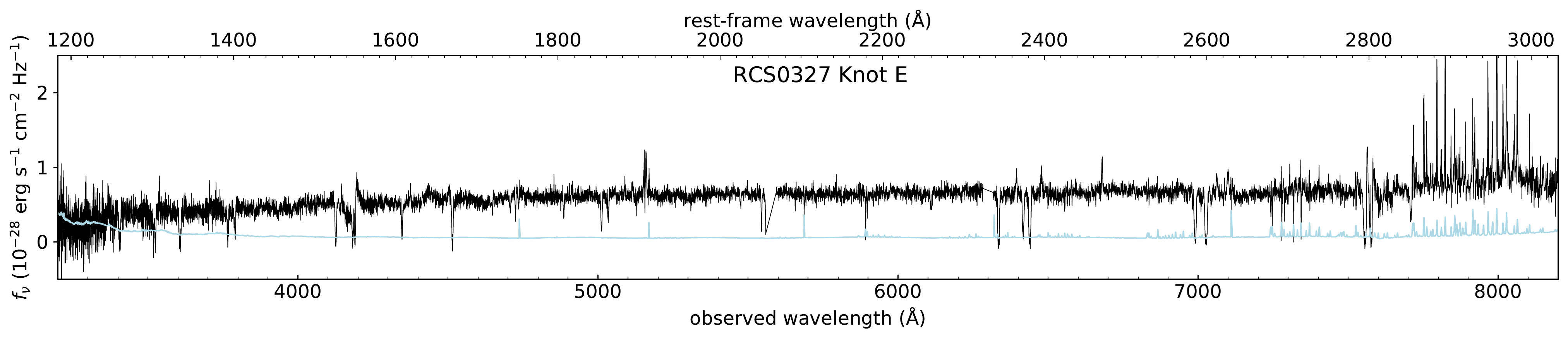} 
\includegraphics[page=2,scale=0.4]{f4.pdf}
\includegraphics[page=3,scale=0.4]{f4.pdf}
\includegraphics[page=4,scale=0.4]{f4.pdf}
\includegraphics[page=5,scale=0.4]{f4.pdf}
\end{figure*}
\begin{figure*}
\includegraphics[page=6,scale=0.4]{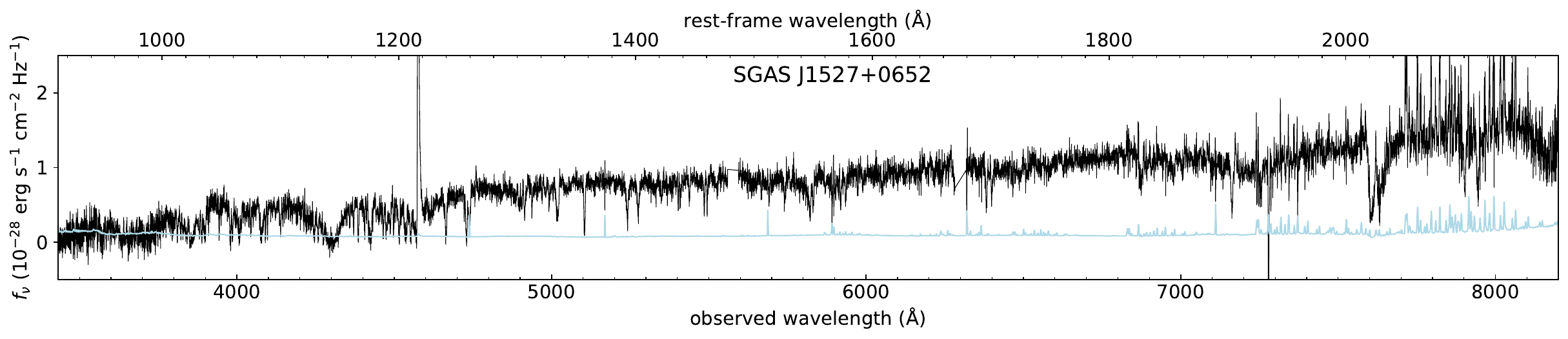}
\includegraphics[page=7,scale=0.4]{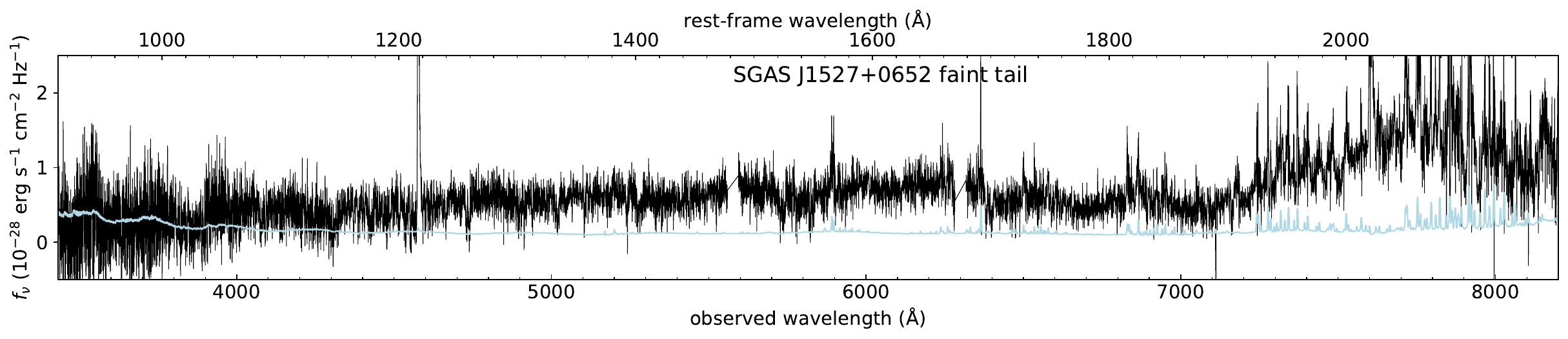}
\includegraphics[page=8,scale=0.4]{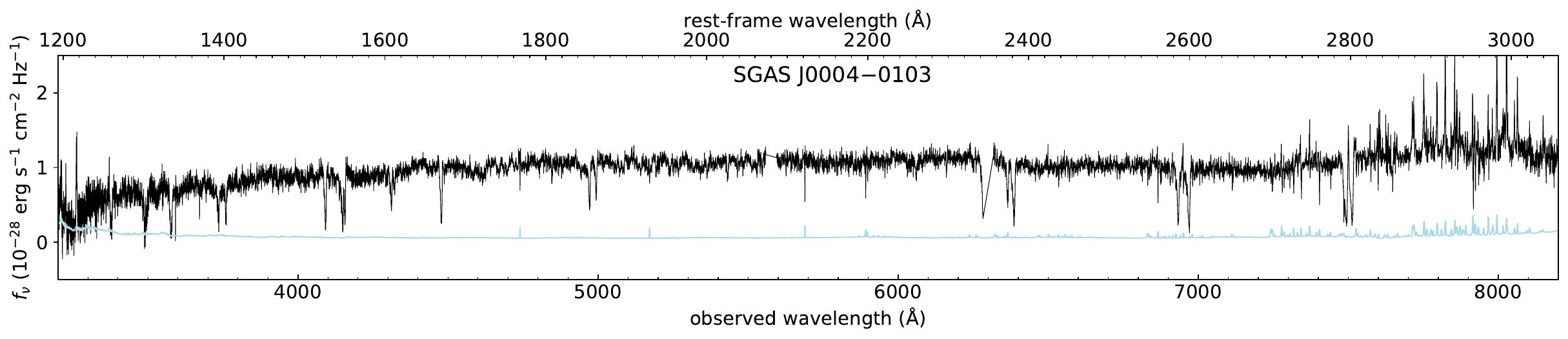}
\includegraphics[page=9,scale=0.4]{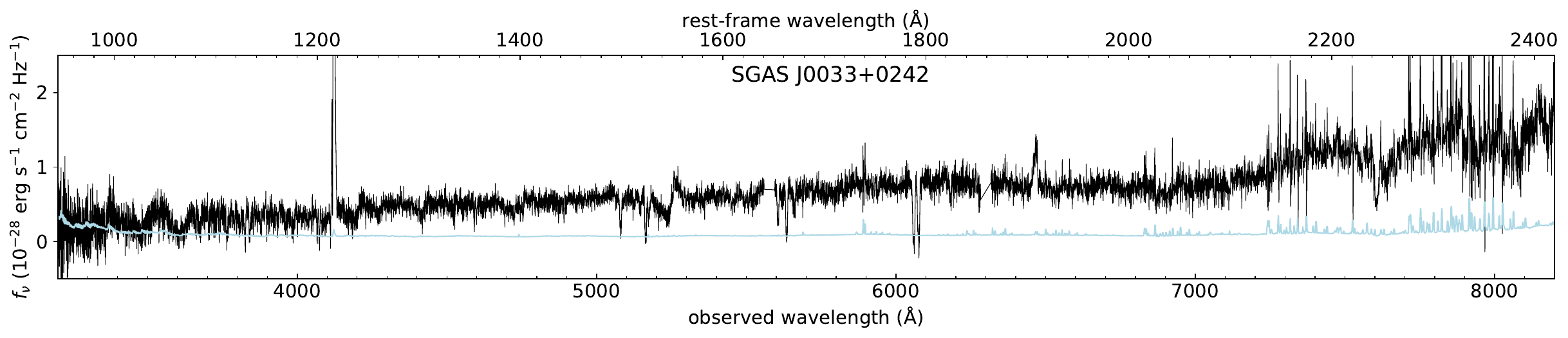}
\includegraphics[page=10,scale=0.4]{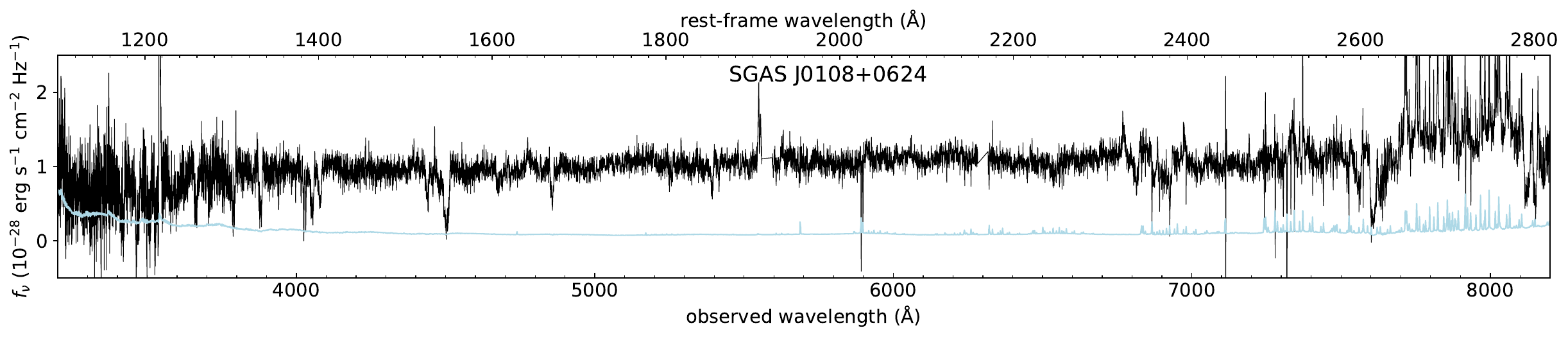} 
\end{figure*}
\begin{figure*}
\includegraphics[page=11,scale=0.4]{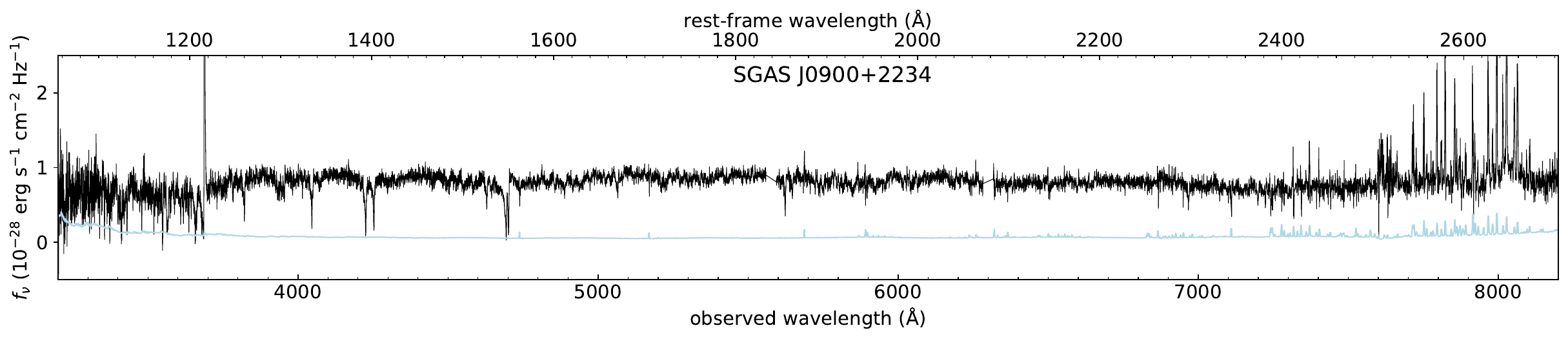} 
\includegraphics[page=12,scale=0.4]{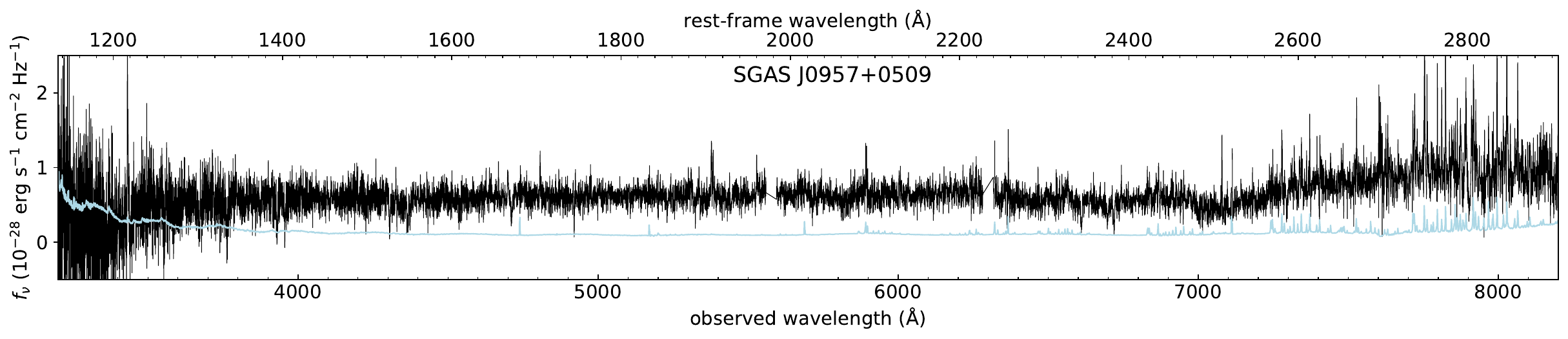} 
\includegraphics[page=13,scale=0.4]{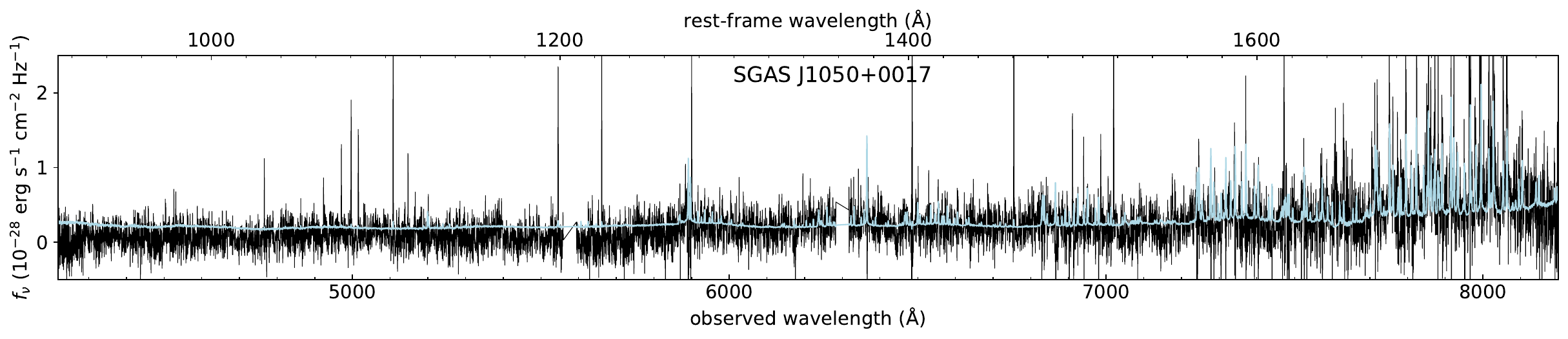} 
\includegraphics[page=14,scale=0.4]{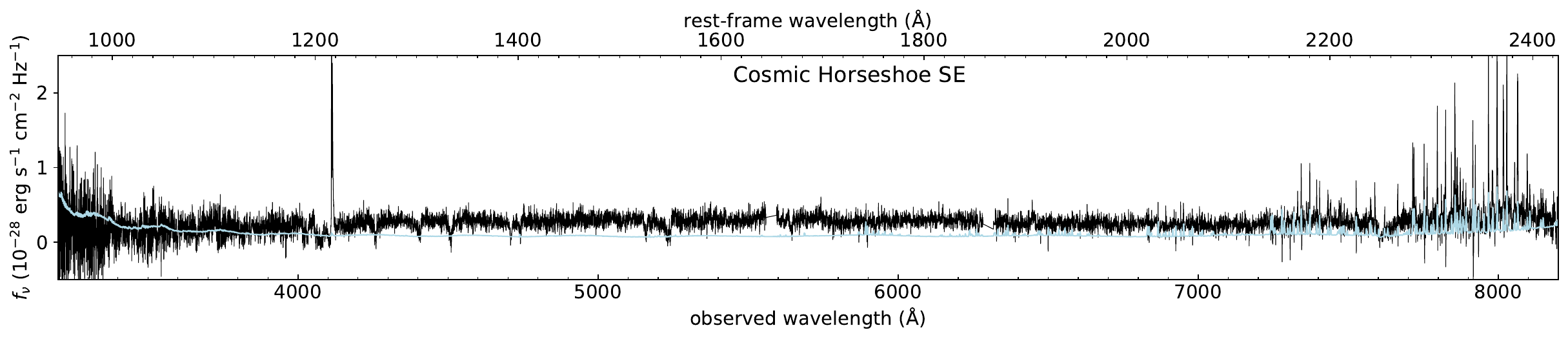} 
\includegraphics[page=15,scale=0.4]{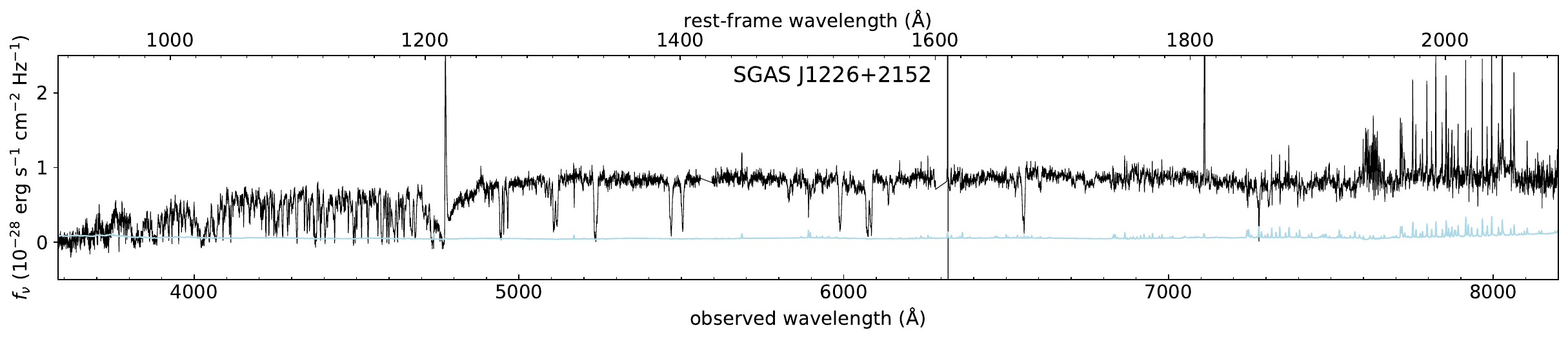} 
\end{figure*}
\begin{figure*}
\includegraphics[page=16,scale=0.4]{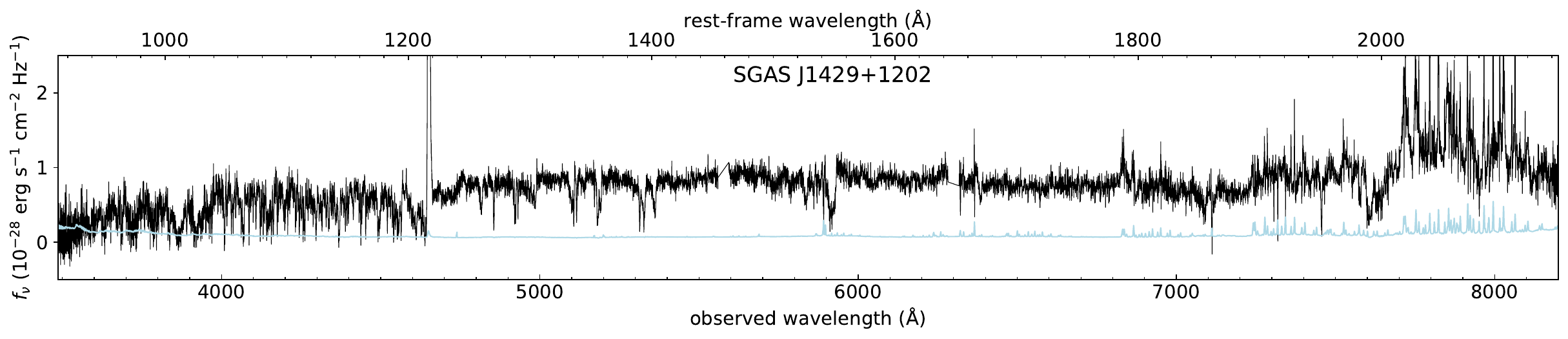} 
\includegraphics[page=17,scale=0.4]{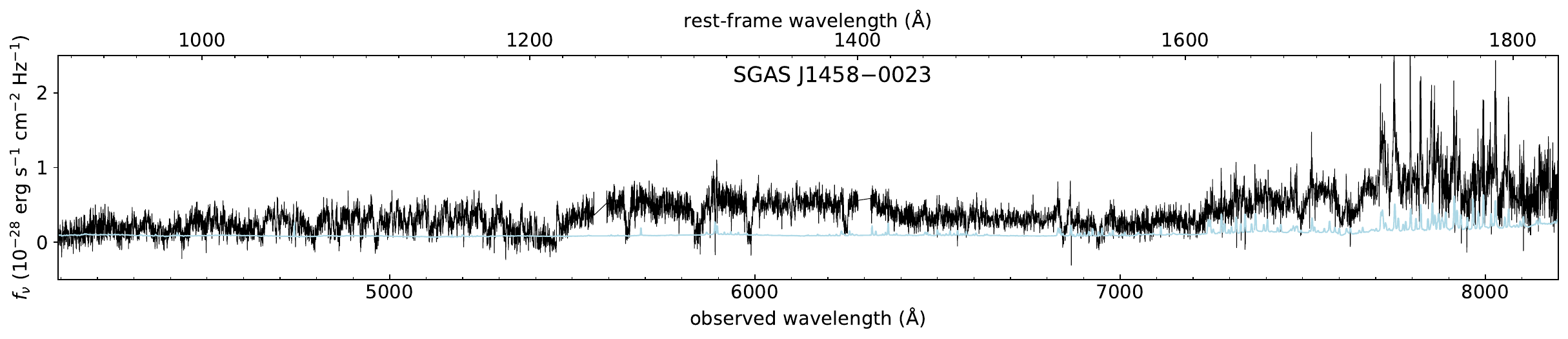} 
\includegraphics[page=18,scale=0.4]{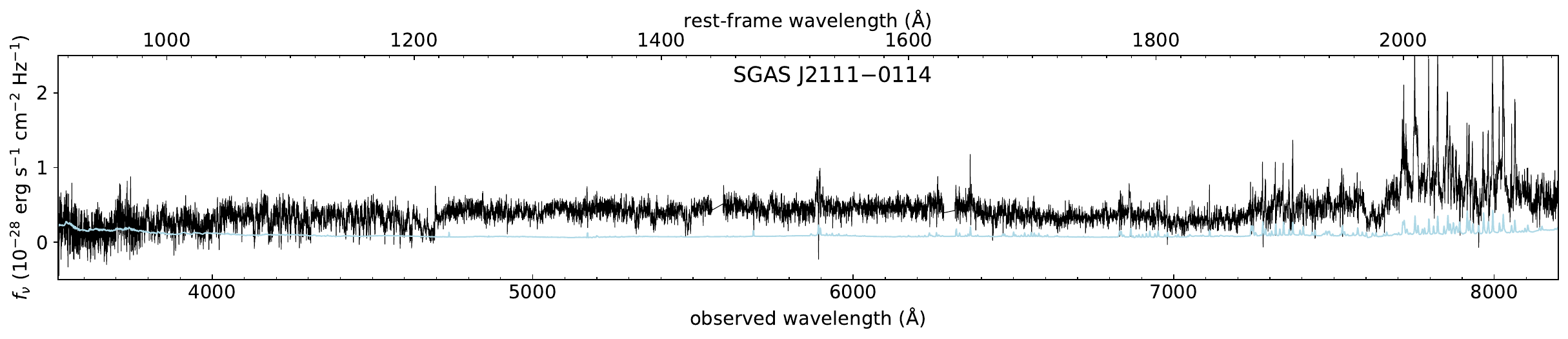} 
\includegraphics[page=19,scale=0.4]{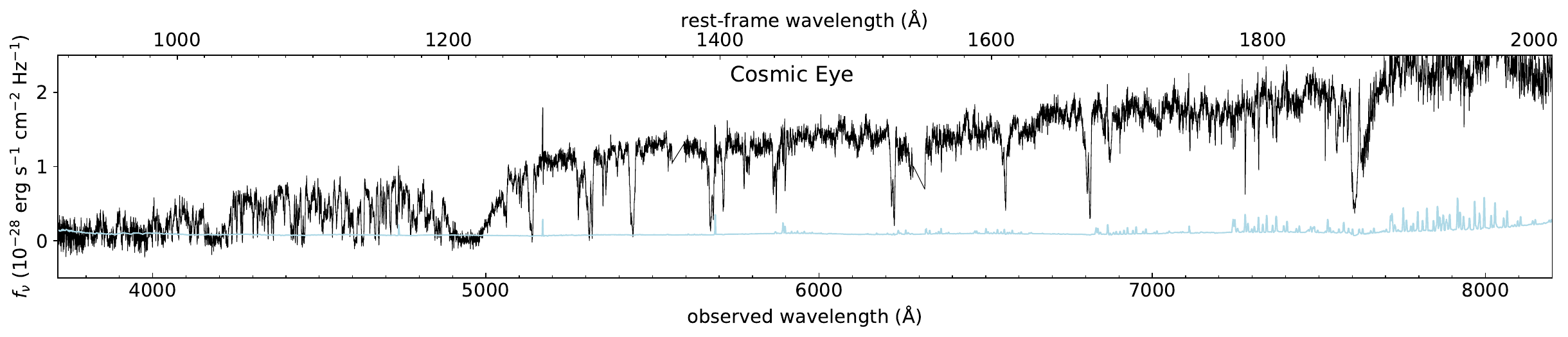} 
\includegraphics[page=20,scale=0.4]{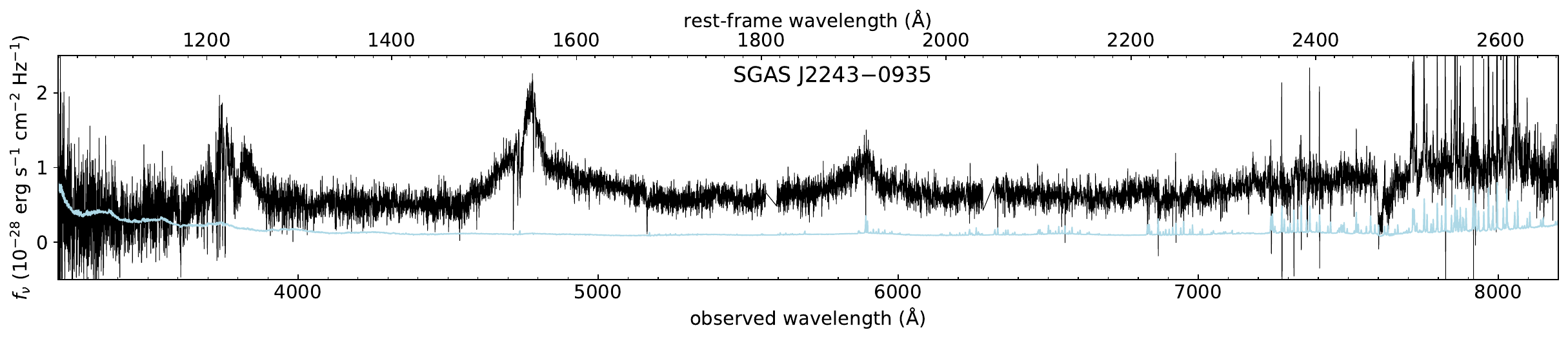} 
\figcaption{The spectra, and their associated $1\sigma$ uncertainty spectra. 
Wavelengths are in vacuum, are barycentric--corrected, and are in units of \AA ;
 the lower x-axis shows observed-frame wavelength, and the upper x-axis
shows rest-frame wavelength.  We have masked the bright sky lines at 5577\AA\ and 6300\AA .
 Flux densities ($f_{\nu}$) are in the observed frame, in units of \cgsfnu , and have been corrected for Milky Way reddening.
\label{fig:spectra}}
\end{figure*}

\begin{figure*}
\includegraphics[width=3in]{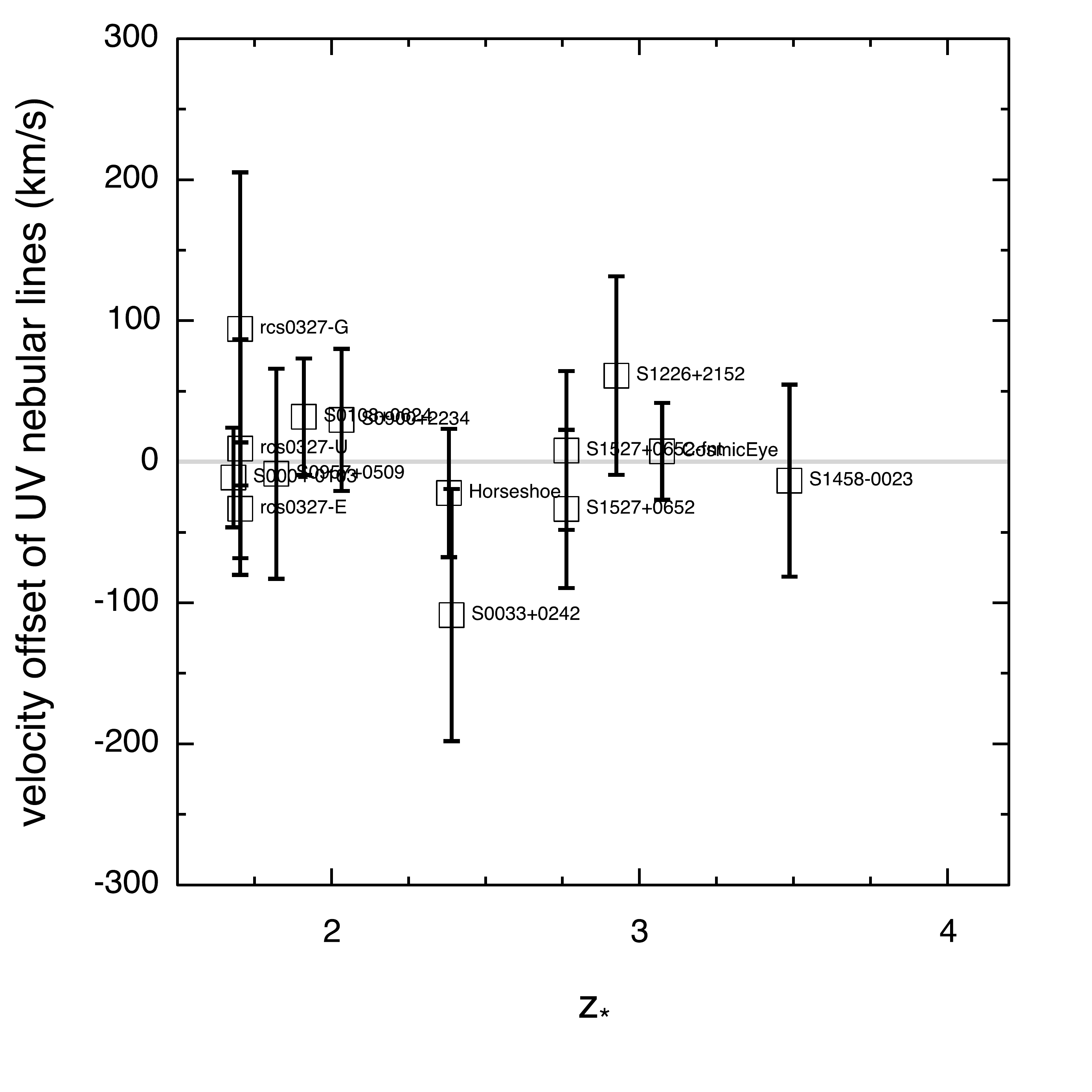} 
\includegraphics[width=3in]{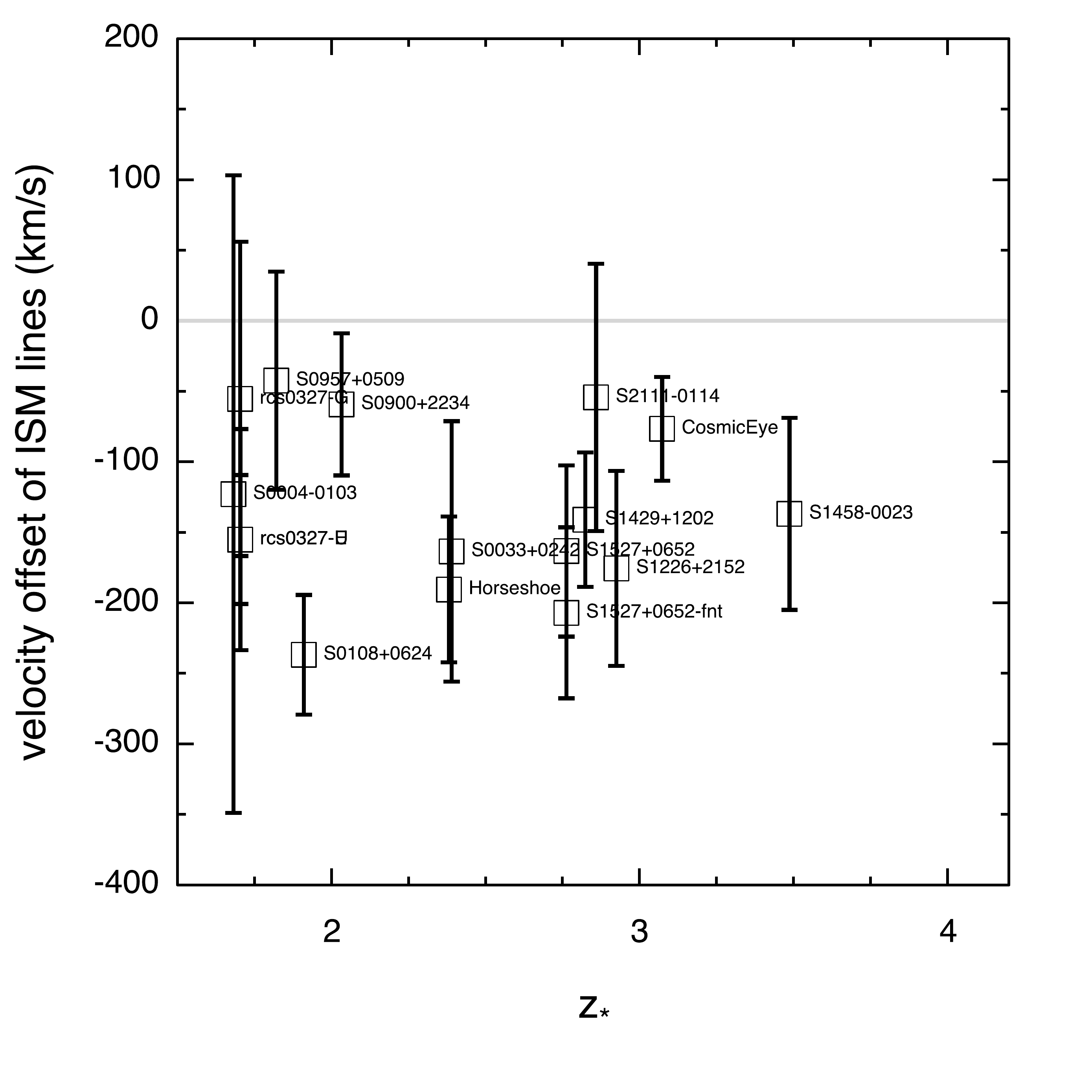} 
\figcaption{Velocity offsets among the stars, nebular gas, and interstellar medium.  
The systemic redshift is taken as the stellar redshift  $z_*$, which is 
measured from Starburst99 fitting to the stellar continuum 
features including the photospheric absorption lines.
The redshift of the ISM lines is measured from the median of Gaussian fits to 
C~II~1334, Si~II~1526, Al~II~1670, Al~III~1854, and Al~III~1862~\AA\ absorption lines.
The redshift of the nebular lines is measured 
from fitting the [C~III]~1907,  C~III] 1909~\AA\ emission line doublet,
except as noted in Table~\ref{tab:redshifts}.
\label{fig:wind_offsets}}
\end{figure*}

\begin{turnpage}
\input{tab_sample_v2}

\input{tab_metallicity_v1.tex}  
\clearpage 
\end{turnpage}
\input{tab_obslog}
\clearpage 
\begin{turnpage}
\input{tab_mage_redshifts_v3.tex}  
\clearpage
\end{turnpage}
\clearpage

\end{document}

%% file: definitions.tex
\newcommand{\megasaura}{M\textsc{eg}a\textsc{S}a\textsc{ura}}
\newcommand{\megasauralong}{The Magellan Evolution of Galaxies Spectroscopic and Ultraviolet Reference Atlas}
\newcommand{\rcsohthree}{RCS-GA~032727$-$132609}
\newcommand{\stwelve}{SGAS J122651.3$+$215220}
\newcommand{\sfifteen}{SGAS~J152745.1$+$065219}

\newcommand{\etal}{et~al.}
\newcommand{\cgsfnu}{erg~s$^{-1}$~cm$^{-2}$~Hz$^{-1}$}

\newcommand{\kms}{\hbox{km~s$^{-1}$}}

\newcommand{\ArII}{\hbox{{\rm Ar}\kern 0.1em{\sc ii}}}
\newcommand{\ArIII}{\hbox{{\rm Ar}\kern 0.1em{\sc iii}}}
\newcommand{\CIV}{\hbox{{\rm C}\kern 0.1em{\sc iv}}}
\newcommand{\HI}{\hbox{{\rm H}\kern 0.1em{\sc i}}}
\newcommand{\HII}{\hbox{{\rm H}\kern 0.1em{\sc ii}}}
\newcommand{\HeI}{\hbox{{\rm He}\kern 0.1em{\sc i}}}
\newcommand{\HeII}{\hbox{{\rm He}\kern 0.1em{\sc ii}}}
\newcommand{\NII}{\hbox{{\rm N}\kern 0.1em{\sc ii}}}
\newcommand{\OI}{\hbox{{\rm O}\kern 0.1em{\sc i}}}
\newcommand{\OII}{\hbox{{\rm O}\kern 0.1em{\sc ii}}}
\newcommand{\OIII}{\hbox{{\rm O}\kern 0.1em{\sc iii}}}
\newcommand{\OIIlong}{{\rm O}\kern 0.1em{\sc ii}~$\lambda 3727$} 
\newcommand{\FeII}{\hbox{{\rm Fe}\kern 0.1em{\sc ii}}}
\newcommand{\NeII}{\hbox{{\rm Ne}\kern 0.1em{\sc ii}}}
\newcommand{\NeIII}{\hbox{{\rm Ne}\kern 0.1em{\sc iii}}}
\newcommand{\NeV}{\hbox{{\rm Ne}\kern 0.1em{\sc v}}}
\newcommand{\SII}{\hbox{{\rm S}\kern 0.1em{\sc ii}}}
\newcommand{\SIII}{\hbox{{\rm S}\kern 0.1em{\sc iii}}}
\newcommand{\SIV}{\hbox{{\rm S}\kern 0.1em{\sc iv}}}
\newcommand{\SiIV}{\hbox{{\rm Si}\kern 0.1em{\sc iv}}}
\newcommand{\MgII}{\hbox{{\rm Mg}\kern 0.1em{\sc ii}}}
\newcommand{\Halpha}{\hbox{{\rm H}\kern 0.1em$\alpha$}}
\newcommand{\Hbeta}{\hbox{{\rm H}\kern 0.1em$\beta$}}
\newcommand{\Heopta}{\hbox{{\rm He}\kern 0.1em{\sc i}}~$6678$}
\newcommand{\Heoptb}{\hbox{{\rm He}\kern 0.1em{\sc i}}~$5876$}
\newcommand{\Heoptc}{\hbox{{\rm He}\kern 0.1em{\sc i}}~$4471$}
\newcommand{\Brgam}{\hbox{{\rm Br}\kern 0.1em$\gamma$}}
\newcommand{\Brten}{\hbox{{\rm Br}\kern 0.1em$10$}}
\newcommand{\Breleven}{\hbox{{\rm Br}\kern 0.1em$11$}}
\newcommand{\HeIh}{\hbox{{\rm He}\kern 0.1em{\sc i}}~$1.7$~{\micron}}
\newcommand{\HeIk}{\hbox{{\rm He}\kern 0.1em{\sc i}}~$2.06$~{\micron}}
\newcommand{\squishlist}{
   \begin{list}{$\bullet$}
    { \setlength{\itemsep}{0pt}      \setlength{\parsep}{1pt}
      \setlength{\topsep}{3pt}       \setlength{\partopsep}{0pt}
      \setlength{\leftmargin}{1.5em} \setlength{\labelwidth}{1em}
      \setlength{\labelsep}{0.5em} } }
\newcommand{\squishend}{
    \end{list}  }

%% file: tab_sample_v2.tex
\begin{deluxetable*}{lllrrlll}
\tabletypesize{\scriptsize}
\tablecolumns{8}
\tablewidth{0pc}
\tablenum{1}
\tablecaption{The \megasaura\ sample. \label{tab:sample}}
\tablehead{\colhead{Source name} &  \colhead{RA (J2000)} &  \colhead{DEC (J2000)} & \colhead{t (hr)} & \colhead{R} & \colhead{Galactic E(B-V)} & \colhead{Discovery} & \colhead{References}}
\startdata
SGAS~J000451.7$-$010321      & 00:04:51.685 & $-$01:03:20.86  & 13.89  & $2750 \pm 100$  &    0.031     & SGAS1   &  \citet{Rigby:2014hq} \\
SGAS~J003341.5$+$024217      & 00:33:41.547 & $+$02:42:16.58  &  7.0   & $2570 \pm 200$  &    0.021     & SGAS2   &   \\
SGAS~J010842.2$+$062444      & 01:08:42.206 & $+$06:24:44.41  &  7.50  & $4380 \pm 200$  &    0.026     & SGAS1   &  \citet{Rigby:2014hq} \\
RCSGA~032727$-$132609        &              &                 &        &                 &                 & \citet{Wuyts:2010gy}   &   \\
~~~Knot~E                    & 03:27:27.975 & $-$13:26:09.00  & 10.00  & $3650 \pm 120$  &    0.079     &    &   \\
~~~Knot~U                    & 03:27:28.256 & $-$13:26:12.90  &  7.94  & $2880 \pm 80$   &              &    &   \\
~~~Knot~B                    & 03:27:27.796 & $-$13:26:07.76  &  2.83  & $3300 \pm 100$  &              &    &   \\
~~~Knot~G                    & 03:27:26.626 & $-$13:26:15.30  &  7.33  & $2830 \pm 140$  &              &    &   \\
~~~counterarc                & 03:27:27.189 & $-$13:26:54.30  &  1.75  & $2600 \pm 100$  &              &    &   \\  
SGAS~J090003.3$+$223408      & 09:00:03.325 & $+$22:34:07.57  & 11.00 &	$2530 \pm 80$    &    0.031     & SGAS1; \citet{Diehl:2009dd}  &   \\
SGAS~J095738.7$+$050929      & 09:57:38.740 & $+$05:09:28.67  &  6.42 &	$3580 \pm 170$   &    0.052     & SGAS1   &  \citet{Bayliss:2011fr} \\
SGAS~J105039.6$+$001730      & 10:50:39.575 & $+$0:17:30.13   &  1.00 &	$4650 \pm 230$   &    0.044     & SGAS1   &  \citet{Oguri:2012bg} \\
Cosmic Horseshoe SE          & 11:48:33.264 & $+$19:29:59.11  &  3.89 & $3980 \pm 220$   &    0.047     & \citet{Belokurov:2007bv}   &   \\
SGAS~J122651.3$+$215220      & 12:26:51.316 & $+$21:52:20.00  & 12.42 &	$4010 \pm 170$   &    0.021     & SGAS1   &  \citet{Koester:2010ky} \\
SGAS~J142954.9$+$120239      & 14:29:54.857 & $+$12:02:38.68  &  9.25 &	$3500 \pm 180$   &    0.032     & SGAS2; \citet{MarquesChaves:2017hs}  &   \\
SGAS~J145836.1$-$002358      & 14:58:36.143 & $-$00:23:58.17  &  8.00 & $2500 \pm 80$    &    0.059     & SGAS2   &  \\
SGAS~J152745.1$+$065219      &              &                 &       &                  &    0.032     & SGAS1   &  \citet{Hennawi:2008it}, \citet{Koester:2010ky} \\
~~~bright                    & 15:27:45.116 & $+$06:52:19.57&  8.42 &  $2740 \pm 70$     &    0.050     &    &   \\
~~~faint tail                & 15:27:45.391 & $+$06:52:18.89&  5.09  & $2520 \pm 70$     &              &    &   \\
SGAS~J211118.9$-$011431      & 21:11:18.946 & $-$01:14:31.44  & 12.11 & $2650 \pm 80$    &    0.084     & SGAS1  &  \citet{Hennawi:2008it}, \citet{Gralla:2011cn} \\
Cosmic Eye                   & 21:35:12.7   & $-$01:01:42.9   &  9.11  & $2530 \pm 75$   &    0.0475    & \citet{Smail:2007bq}   &   \\ 
SGAS~J224324.2$-$093508      & 22:43:24.221 & $-$09:35:08.55  &  4.00 & $4760 \pm 200$   &    0.048     & SGAS1   & \citet{Bayliss:2011fr}\\
\enddata
\tablecomments{Columns: 
1) Source name.
2) and 3):  Right ascension and declination (J2000) for the center of the MagE slit.  
Astrometry for most targets is from \textit{HST} images; the exceptions are 
SGAS~J142954.9$+$120239 and SGAS~J145836.1$-$002358 where the astrometry is from 
the Nordic Optical Telescope (NOT) images corrected to the USNO system, 
and SGAS~J003341.5$+$024217 where the astrometry is from IMACS/Magellan images 
corrected to the USNO system.
4): Total MagE/Magellan on-source integration time.
5): MagE spectral resolution, defined as $R \equiv \lambda / (FWHM)$, where FWHM is the full width 
at half maximum, measured as the median ($\pm$ the median absolute deviation) of 
measurements of isolated night-sky emission lines in each combined spectrum. 
6) E(B-V) reddening due to the Milky Way Galaxy, from \citet{Green:2015cf}. 
6): Discovery paper(s) for the arc.  We list multiple discovery papers where the arc
was independently found by multiple searches.
7) References: Additional references.  
}
\end{deluxetable*}

%% file: tab_metallicity_v1.tex
\begin{deluxetable}{llllll}
\tabletypesize{\scriptsize}
\tablecolumns{6}
\tablewidth{0pc}
\tablenum{2}
\label{tab:metallicities}
\tablecaption{Oxygen abundances for the \megasaura\ sample.}
\tablehead{\colhead{object}& \colhead{$\log([N II]/H\alpha)$} & \colhead{$12 + \log(O/H)$} & \colhead{method}         & \colhead{notes}                  & \colhead{reference}}
\startdata
SGAS~J000451.7$-$010321   &  $<-1.4$                        & $<8.09$            &   \citet{Pettini:2004bq} linear  &                                    & see Note below\\
SGAS~J003341.5$+$024217   &   & & & & \\
SGAS~J010842.2$+$062444   &   & & & & \\
RCSGA~032727$-$132609     &   $-1.19 \pm 0.07$              & $8.22 \pm 0.04$    & \citet{Pettini:2004bq} linear    &                                    & \citet{Rigby:2011il} \\
SGAS~J090003.3$+$223408   &                                 & $8.18 \pm 0.14$    & \citet{Pettini:2004bq} linear    & weighted average of 2 measurements & \citet{Bian:2010bl} \\
SGAS~J095738.7$+$050929   &   & & & & \\
SGAS~J105039.6$+$001730   &                                 & $8.3 \pm 0.1$      & multiple $R_{23}$ methods        &                                    & \citet{Bayliss:2014ib}\\ 
Cosmic Horseshoe SE       &   $-0.79 \pm 0.04$              & $8.45 \pm 0.03$    &  \citet{Pettini:2004bq} linear   & their ``Aperture 2''               & \citet{Hainline:2009fg}\\
SGAS~J122651.3$+$215220   &   & & & & \\ 
SGAS~J142954.9$+$120239   &   & & & & \\
SGAS~J145836.1$-$002358   &   & & & & \\ 
SGAS~J152745.1$+$065219   &   $< - 0.65$                    & $<8.5$              &  \citet{Pettini:2004bq}         &                                    &  \citet{Wuyts:2012ej}\\   
SGAS~J211118.9$-$011431   &   & & & & \\
Cosmic Eye                &                                 & $8.6$               & $R_{23}$, \citet{Pettini:2004bq, Pilyugin:2005kn}   &  assumed upper branch & \citet{Stark:2008hx}\\
                          &                                 & $7.97^{+0.32}_{-0.23}$ & \citet{Maiolino:2008gs}        &                                    & \citet{Troncoso:2014kg}\\
SGAS~J224324.2$-$093508   &   & & & & \\ 
\enddata
\tablecomments{Complilation of measured oxygen abundances for the \megasaura\ galaxies.  
We list the measured $\log([N II] 6583 /H\alpha)$ ratio, and the resulting inferred 
oxygen abundance (commonly called the ``metallicity'') in the format $12 + \log(O/H)$.  
The measurement for  SGAS~J0004 is from unpublished FIRE/Magellan spectra, 
obtained UT~2010-10-14 and UT~2013-09-13, with a total integration time of 4.18~hr.
}
\end{deluxetable}

%% file: tab_obslog.tex
\LongTables
\begin{deluxetable*}{lllllll}
\tabletypesize{\small}
\tablecolumns{7}
\tablewidth{0pc}
\tablenum{3}
\tablecaption{Observation log. \label{tab:obslog}}
\tablehead{
\colhead{UT Date} &  \colhead{UT Time} &  \colhead{t$_{int}$ (s)} &  
\colhead{sec(z)} & \colhead{slit width (\arcsec)} & 
\colhead{slit angle (\degr)} &  \colhead{par. angle (\degr)}}
\startdata
\cutinhead{SGAS~J000451.7$-$010321}
2010-12-09  & 00:48:47 & $1800$ & $1.189$ & $2.0$ & $125$ & $150$  \\
2010-12-09  & 01:19:26 & $1800$ & $1.254$ & $2.0$ & $125$ & $141$  \\
2013-08-10  & 05:16:34 & $3600$ & $1.360$ & $2.0$ & $222$ & $226$  \\
2013-08-10  & 06:23:21 & $3600$ & $1.184$ & $2.0$ & $200$ & $209$  \\
2013-08-10  & 07:28:20 & $3000$ & $1.132$ & $2.0$ & $170$ & $182$  \\
2013-10-06  & 02:03:58 & $3400$ & $1.258$ & $1.5$ & $212$ & $219$  \\
2013-10-06  & 03:04:36 & $3400$ & $1.152$ & $1.5$ & $187$ & $200$  \\
2013-10-07  & 01:08:30 & $3200$ & $1.450$ & $1.5$ & $225$ & $229$  \\
2013-10-07  & 02:06:07 & $3200$ & $1.242$ & $1.5$ & $210$ & $218$  \\
2013-10-07  & 03:02:59 & $3200$ & $1.150$ & $1.5$ & $186$ & $199$  \\
2015-11-06  & 00:54:30 & $1800$ & $1.163$ & $2.0$ & $302$ & $204$  \\
2015-11-06  & 01:42:00 & $1800$ & $1.132$ & $2.0$ & $125$ & $183$  \\
2015-11-06  & 02:12:47 & $3600$ & $1.139$ & $2.0$ & $125$ & $169$  \\
2015-11-06  & 03:13:28 & $3600$ & $1.215$ & $2.0$ & $125$ & $146$  \\
2015-11-06  & 04:14:17 & $3600$ & $1.408$ & $2.0$ & $125$ & $132$  \\
2015-11-07  & 00:15:51 & $2702$ & $1.220$ & $2.0$ & $208$ & $215$  \\
2015-11-07  & 01:01:38 & $2700$ & $1.151$ & $2.0$ & $208$ & $199$  \\
%
\cutinhead{SGAS~J003341.5$+$024217 }  
2015-11-07  & 01:50:28 & $3600$ & $1.181$ & $2.0$ & $178$ & $189$  \\
2015-11-07  & 02:51:13 & $3600$ & $1.193$ & $2.0$ & $178$ & $165$  \\
2015-11-07  & 03:54:35 & $3600$ & $1.306$ & $2.0$ & $138$ & $144$  \\
2015-11-10  & 00:17:12 & $3600$ & $1.311$ & $2.0$ & $180$ & $216$  \\
2015-11-10  & 01:21:19 & $3600$ & $1.193$ & $2.0$ & $185$ & $196$  \\
2015-11-10  & 02:23:34 & $3600$ & $1.181$ & $2.0$ & $160$ & $171$  \\
2015-11-10  & 03:29:44 & $1800$ & $1.273$ & $2.0$ & $141$ & $148$  \\
2015-11-10  & 04:00:49 & $1800$ & $1.364$ & $2.0$ & $141$ & $140$  \\
\cutinhead{SGAS~J010842.2$+$062444}
2012-08-19  & 06:29:45 & $3600$ & $1.334$ & $1.0$ & $203$ & $210$  \\
2012-08-19  & 07:32:33 & $3600$ & $1.236$ & $1.0$ & $180$ & $190$  \\
2012-08-19  & 08:41:48 & $3600$ & $1.250$ & $1.0$ & $156$ & $165$  \\
2012-08-20  & 08:48:26 & $3600$ & $1.263$ & $1.0$ & $153$ & $161$  \\
2012-09-14  & 07:40:24 & $2400$ & $1.319$ & $0.7$ & $179$ & $152$  \\
2012-09-14  & 08:21:06 & $2400$ & $1.449$ & $0.7$ & $179$ & $142$  \\
2012-09-14  & 09:01:34 & $2100$ & $1.668$ & $0.7$ & $179$ & $134$  \\
2013-08-10  & 08:26:59 & $2700$ & $1.23$ & $2.0$ & $175$ & $184$  \\
2013-08-10  & 09:15:50 & $3000$ & $1.25$ & $2.0$ & $156$ & $165$  \\
\cutinhead{RCSGA 032727-132609 knot E}
2008-07-31  & 09:03:00 & $1800$ & $1.268$ & $2.0$ & $240$ & $240$  \\
2010-12-09  & 02:18:04 & $2700$ & $1.053$ & $2.0$ & $196$ & $210$  \\
2010-12-09  & 03:09:56 & $3600$ & $1.039$ & $2.0$ & $164$ & $172$  \\
2010-12-09  & 04:20:47 & $3600$ & $1.099$ & $2.0$ & $152$ & $133$  \\
2010-12-09  & 05:29:20 & $3600$ & $1.271$ & $2.0$ & $88$ & $120$  \\
2010-12-09  & 06:30:52 & $2700$ & $1.601$ & $2.0$ & $88$ & $117$  \\
2010-12-10  & 00:48:00 & $3600$ & $1.198$ & $1.0$ & $235$ & $237$  \\
2010-12-10  & 01:59:40 & $3600$ & $1.066$ & $1.0$ & $207$ & $218$  \\
2010-12-10  & 03:08:19 & $3600$ & $1.040$ & $1.0$ & $183$ & $170$  \\
2010-12-10  & 04:26:51 & $3600$ & $1.116$ & $1.0$ & $150$ & $130$  \\
2010-12-10  & 05:32:13 & $3600$ & $1.297$ & $1.0$ & $93$ & $120$  \\
\cutinhead{RCSGA 032727-132609  knot U}
2008-07-31  & 09:34:55 & $1800$ & $1.171$ & $2.0$ & $240$ & $236$  \\
2008-07-31  & 10:05:46 & $900$ & $1.107$ & $2.0$ & $240$ & $229$  \\
2010-02-16  & 00:39:22 & $1800$ & $1.213$ & $2.0$ & $30$ & $122$  \\
2010-02-16  & 01:10:14 & $1800$ & $1.323$ & $2.0$ & $30$ & $119$  \\
2010-02-17  & 00:49:53 & $1800$ & $1.260$ & $2.0$ & $103$ & $121$  \\
2010-02-17  & 01:24:22 & $1800$ & $1.409$ & $2.0$ & $103$ & $118$  \\
2010-02-17  & 01:55:04 & $1800$ & $1.604$ & $2.0$ & $103$ & $117$  \\
2013-10-06  & 05:07:01 & $3300$ & $1.185$ & $1.5$ & $231$ & $236$  \\
2013-10-06  & 06:07:41 & $3300$ & $1.073$ & $1.5$ & $157$ & $220$  \\
2013-10-06  & 07:06:41 & $3300$ & $1.038$ & $1.5$ & $157$ & $183$  \\
2013-10-07  & 07:04:20 & $3500$ & $1.038$ & $1.5$ & $161$ & $182$  \\
2013-10-07  & 08:06:37 & $3500$ & $1.071$ & $1.5$ & $131$ & $141$  \\
\cutinhead {RCSGA 032727-132609 knot B}
2013-10-06  & 04:07:25 & $3300$ & $1.414$ & $1.5$ & $240$ & $242$  \\
2013-10-07  & 04:01:37 & $3600$ & $1.424$ & $1.5$ & $241$ & $242$  \\
2013-10-07  & 05:05:37 & $3300$ & $1.178$ & $1.5$ & $231$ & $236$  \\
\cutinhead{RCSGA 032727-132609  knot G}
2013-10-07  & 06:04:28 & $3300$ & $1.072$ & $1.5$ & $199$ & $220$  \\
2013-10-06  & 08:07:24 & $3300$ & $1.068$ & $1.5$ & $199$ & $142$  \\
2015-11-06  & 05:30:07 & $3600$ & $1.042$ & $2.0$ & $199$ & $165$  \\
2015-11-06  & 06:33:59 & $3600$ & $1.103$ & $2.0$ & $199$ & $132$  \\
2015-11-07  & 05:00:52 & $3600$ & $1.038$ & $2.0$ & $199$ & $185$  \\
2015-11-07  & 06:01:31 & $3600$ & $1.066$ & $2.0$ & $199$ & $143$  \\
2015-11-10  & 04:42:00 & $1800$ & $1.039$ & $2.0$ & $199$ & $190$  \\
2015-11-10  & 05:13:36 & $3600$ & $1.041$ & $2.0$ & $199$ & $165$  \\
\cutinhead{RCSGA 032727-132609 counterarc}
2015-11-10  & 06:40:52 & $3600$ & $1.146$ & $2.0$ & $270$ & $307$  \\
2015-11-10  & 07:42:52 & $2701$ & $1.347$ & $2.0$ & $270$ & $299$  \\
\cutinhead{SGAS~090003.3$+$2234:08}
2009-04-21  & 23:26:06 & $1200$ & $1.613$ & $2.0$ & $182$ & $185$  \\
2009-04-21  & 23:49:13 & $1800$ & $1.609$ & $2.0$ & $182$ & $178$  \\
2009-04-22  & 00:19:54 & $1800$ & $1.637$ & $2.0$ & $182$ & $170$  \\
2009-04-22  & 00:55:55 & $1800$ & $1.723$ & $2.0$ & $154$ & $160$  \\
2009-04-22  & 23:14:37 & $1800$ & $1.619$ & $2.0$ & $183$ & $187$  \\
2009-04-22  & 23:51:25 & $1800$ & $1.611$ & $2.0$ & $183$ & $176$  \\
2009-04-23  & 00:24:58 & $1800$ & $1.652$ & $2.0$ & $160$ & $167$  \\
2009-04-23  & 00:55:41 & $1800$ & $1.736$ & $2.0$ & $160$ & $159$  \\
2009-04-23  & 23:12:48 & $1800$ & $1.617$ & $2.0$ & $184$ & $186$  \\
2009-04-23  & 23:45:24 & $3600$ & $1.610$ & $2.0$ & $184$ & $177$  \\
2009-04-27  & 00:45:54 & $2400$ & $1.758$ & $2.0$ & $153$ & $158$  \\
2010-02-16  & 03:07:32 & $1800$ & $1.661$ & $2.0$ & $189$ & $194$  \\
2010-02-16  & 03:38:23 & $1800$ & $1.616$ & $2.0$ & $189$ & $186$  \\
2010-02-16  & 04:19:03 & $1800$ & $1.616$ & $2.0$ & $168$ & $174$  \\
2010-02-16  & 04:49:44 & $1800$ & $1.661$ & $2.0$ & $168$ & $166$  \\
2010-02-16  & 05:20:28 & $1800$ & $1.751$ & $2.0$ & $168$ & $158$  \\
2010-02-17  & 02:32:03 & $1800$ & $1.754$ & $2.0$ & $196$ & $202$  \\
2010-02-17  & 03:03:01 & $1800$ & $1.662$ & $2.0$ & $196$ & $194$  \\
2010-02-17  & 03:40:20 & $3600$ & $1.612$ & $2.0$ & $177$ & $184$  \\
2010-02-17  & 04:44:19 & $1800$ & $1.658$ & $2.0$ & $159$ & $167$  \\
\cutinhead{SGAS~J095738.7$+$050929}
2010-12-09  & 07:44:11 & $2100$ & $1.353$ & $2.0$ & $209$ & $215$  \\
2010-12-10  & 06:48:36 & $2700$ & $1.579$ & $1.0$ & $221$ & $225$  \\
2010-12-10  & 07:35:19 & $2700$ & $1.369$ & $1.0$ & $221$ & $216$  \\
2014-05-29  & 23:23:51 & $3600$ & $1.276$ & $2.0$ & $147$ & $154$  \\
2014-05-31  & 00:01:46 & $2400$ & $1.388$ & $2.0$ & $125$ & $143$  \\
2015-06-09  & 23:08:45 & 2400 & 1.341 & 1.0 & 143 & 146\\
2015-06-09  & 23:50:18 & 2400 & 1.502 & 1.0 & 143 & 137\\
2015-06-13  & 23:14:09 & 2400 & 1.412 & 2.0 & 139 & 141\\
2015-06-14  & 00:00:31 & 2400 & 1.655 & 2.0 & 129 & 133\\
\cutinhead{SGAS~J105039.6$+$001730}
2013-05-06  & 02:30:28 & $3600$ & $1.303$ & $1.0$ & $136$ & $140$  \\
\cutinhead{Cosmic Horseshoe SE}
2008-02-06  & 06:22:20 & $1800$ & $1.589$ & $0.85$ & $191$ & $199$  \\
2008-02-06  & 06:53:11 & $1800$ & $1.530$ & $0.85$ & $191$ & $190$  \\
2008-02-06  & 07:24:51 & $1400$ & $1.58$ & $0.85$ & $191$ & $181$  \\
2008-02-08  & 05:59:01 & $1800$ & $1.635$ & $2.0$ & $196$ & $203$  \\
2008-02-08  & 06:29:37 & $1800$ & $1.555$ & $2.0$ & $196$ & $195$  \\
2008-02-08  & 07:04:29 & $1800$ & $1.513$ & $2.0$ & $177$ & $185$  \\
2008-02-08  & 07:35:03 & $1200$ & $1.512$ & $2.0$ & $177$ & $176$  \\
2008-02-09  & 05:30:20 & $2400$ & $1.735$ & $2.0$ & $202$ & $209$  \\
\cutinhead{SGAS~J122651.3$+$215220}
2008-02-09  & 06:15:16 & $1800$ & $1.797$ & $2.0$ & $200$ & $206$  \\
2008-02-09  & 06:45:41 & $1800$ & $1.680$ & $2.0$ & $200$ & $199$  \\
2008-02-09  & 07:18:35 & $1800$ & $1.609$ & $2.0$ & $200$ & $190$  \\
2008-02-09  & 07:49:04 & $1800$ & $1.584$ & $2.0$ & $200$ & $182$  \\
2008-02-09  & 08:21:29 & $1800$ & $1.598$ & $2.0$ & $165$ & $172$  \\
2008-02-09  & 08:51:56 & $1505$ & $1.650$ & $2.0$ & $165$ & $164$  \\
2009-04-22  & 02:57:39 & $3600$ & $1.586$ & $2.0$ & $176$ & $183$  \\
2009-04-23  & 02:21:32 & $3000$ & $1.620$ & $2.0$ & $185$ & $192$  \\
2009-04-23  & 03:17:10 & $3000$ & $1.587$ & $2.0$ & $169$ & $177$  \\
2009-04-23  & 04:07:56 & $1800$ & $1.666$ & $2.0$ & $169$ & $163$  \\
2010-02-16  & 06:01:20 & $1800$ & $1.730$ & $0.7$ & $196$ & $203$  \\
2010-02-16  & 06:32:07 & $1800$ & $1.639$ & $0.7$ & $196$ & $195$  \\
2010-02-16  & 07:05:56 & $3600$ & $1.590$ & $0.7$ & $177$ & $186$  \\
2010-02-16  & 08:11:06 & $3000$ & $1.626$ & $0.7$ & $160$ & $167$  \\
2010-02-17  & 05:32:21 & $3600$ & $1.845$ & $0.7$ & $202$ & $209$  \\
2010-02-17  & 06:37:11 & $3600$ & $1.621$ & $0.7$ & $185$ & $192$  \\
2010-02-17  & 07:43:53 & $3600$ & $1.594$ & $0.7$ & $166$ & $174$  \\
2010-02-17  & 08:44:36 & $1800$ & $1.731$ & $0.7$ & $166$ & $158$  \\
\cutinhead{SGAS~J142954.9$+$120239}
2013-05-04  & 03:13:58 & $3600$ & $1.399$ & $1.0$ & $195$ & $202$  \\
2013-05-04  & 04:19:51 & $3600$ & $1.326$ & $1.0$ & $173$ & $182$  \\
2013-05-06  & 03:41:45 & $3600$ & $1.342$ & $1.0$ & $183$ & $191$  \\
2013-05-06  & 04:45:15 & $4500$ & $1.338$ & $1.0$ & $162$ & $171$  \\
2014-05-30  & 00:31:40 & $3600$ & $1.617$ & $2.0$ & $212$ & $217$  \\
2014-05-30  & 01:36:12 & $3600$ & $1.391$ & $2.0$ & $193$ & $201$  \\
2014-05-31  & 00:45:13 & $3600$ & $1.535$ & $2.0$ & $207$ & $213$  \\
2014-05-31  & 01:47:00 & $3600$ & $1.365$ & $2.0$ & $188$ & $197$  \\
2014-05-31  & 02:49:03 & $3600$ & $1.327$ & $2.0$ & $167$ & $177$  \\
\cutinhead{SGAS~J145836.1$-$002358}
2013-08-09  & 23:38:43 & $2700$ & $1.20$ & $2.0$ & $143$ & $151$  \\
2013-08-10  & 00:29:01 & $2700$ & $1.32$ & $2.0$ & $132$ & $137$  \\
2013-08-10  & 23:32:16 & $2700$ & $1.19$ & $2.0$ & $143$ & $152$  \\
2013-08-11  & 00:20:52 & $2700$ & $1.31$ & $2.0$ & $133$ & $138$  \\
2014-05-30  & 02:40:12 & $3600$ & $1.149$ & $2.0$ & $182$ & $194$  \\
2014-05-30  & 03:44:19 & $3600$ & $1.151$ & $2.0$ & $155$ & $166$  \\
2014-05-30  & 04:46:18 & $3600$ & $1.245$ & $2.0$ & $137$ & $144$  \\
2014-05-31  & 03:53:38 & $3600$ & $1.163$ & $2.0$ & $150$ & $160$  \\
2014-05-31  & 04:55:43 & $3600$ & $1.280$ & $2.0$ & $134$ & $140$  \\
\cutinhead{SGAS J152745.1$+$065219 bright}
2008-02-08  & 08:49:42 & $900$ & $1.479$ & $2.0$ & $213$ & $219$  \\
2008-07-30  & 01:23:08 & $1800$ & $1.386$ & $2.0$ & $282$ & $326$  \\
2008-07-30  & 01:55:14 & $1800$ & $1.515$ & $2.0$ & $282$ & $319$  \\
2008-07-30  & 23:27:11 & $1800$ & $1.235$ & $2.0$ & $177$ & $183$  \\
2008-07-30  & 23:59:37 & $1800$ & $1.242$ & $2.0$ & $177$ & $171$  \\
2008-07-31  & 23:52:11 & $3600$ & $1.240$ & $2.0$ & $163$ & $172$  \\
2008-08-01  & 01:00:23 & $3600$ & $1.343$ & $2.0$ & $144$ & $150$  \\
2008-08-01  & 02:06:54 & $1812$ & $1.624$ & $2.0$ & $132$ & $136$  \\
2009-04-22  & 05:46:55 & $3600$ & $1.240$ & $2.0$ & $180$ & $188$  \\
2009-04-22  & 06:47:31 & $1800$ & $1.253$ & $2.0$ & $180$ & $166$  \\
2015-06-10  & 00:39:05 & 3900   &  1.525  & 1.0   & 216  & 222\\
2015-06-10  & 01:44:28 & 3900   & 1.305   & 1.0   & 216  & 206\\
\cutinhead{SGAS J152745.1$+$065219 faint tail}
2015-06-13 & 01:10:54 & 3000 & 1.358 &  2.0  & 207 &  212\\
2015-06-13 & 02:05:59 & 3000 & 1.254 &  2.0  & 186 &  195\\
2015-06-13 & 03:13:17 & 2940 & 1.243 &  2.0  & 162 &  170\\
2015-06-14 & 00:45:52 & 3600 & 1.427 &  2.0  & 210 &  217\\
2015-06-14 & 01:46:58 & 3000 & 1.273 &  2.0  & 210 &  200\\
2015-06-14 & 02:38:03 & 2800 & 1.234 &  2.0  & 210 &  182\\
\cutinhead{SGAS~J211118.9$-$011431}
2013-08-10  & 01:36:17 & $3600$ & $1.610$ & $2.0$ & $230$ & $233$  \\
2013-08-10  & 02:40:16 & $3600$ & $1.300$ & $2.0$ & $216$ & $223$  \\
2013-08-10  & 03:46:00 & $4500$ & $1.160$ & $2.0$ & $193$ & $203$  \\
2013-10-05  & 23:56:07 & $3600$ & $1.167$ & $1.5$ & $197$ & $206$  \\
2013-10-06  & 00:59:14 & $3600$ & $1.130$ & $1.5$ & $165$ & $178$  \\
2014-05-30  & 06:05:56 & $3600$ & $1.721$ & $2.0$ & $232$ & $235$  \\
2014-05-30  & 07:08:42 & $3600$ & $1.353$ & $2.0$ & $220$ & $226$  \\
2014-05-30  & 08:11:30 & $3600$ & $1.186$ & $2.0$ & $199$ & $210$  \\
2014-05-30  & 09:13:58 & $1300$ & $1.131$ & $2.0$ & $171$ & $185$  \\
2014-05-31  & 06:27:38 & $3600$ & $1.535$ & $2.0$ & $228$ & $232$  \\
2014-05-31  & 07:29:40 & $3600$ & $1.270$ & $2.0$ & $213$ & $221$  \\
2014-05-31  & 08:31:26 & $2700$ & $1.154$ & $2.0$ & $189$ & $201$  \\
2014-05-31  & 09:18:02 & $2700$ & $1.130$ & $2.0$ & $167$ & $181$  \\
\cutinhead{Cosmic Eye}
2008-07-28  & 03:04:54 & $1800$ & $1.520$ & $2.0$ & $229$ & $231$  \\
2008-07-28  & 03:36:57 & $1800$ & $1.363$ & $2.0$ & $229$ & $226$  \\
2008-07-28  & 06:19:28 & $1600$ & $1.140$ & $2.0$ & $158$ & $168$  \\
2008-07-28  & 06:48:40 & $1800$ & $1.166$ & $2.0$ & $158$ & $156$  \\
2008-07-28  & 07:25:19 & $1800$ & $1.230$ & $2.0$ & $137$ & $144$  \\
2008-07-28  & 07:56:00 & $1800$ & $1.317$ & $2.0$ & $137$ & $137$  \\
2008-07-31  & 03:24:03 & $1800$ & $1.368$ & $2.0$ & $224$ & $226$  \\
2008-07-31  & 03:55:23 & $1800$ & $1.263$ & $2.0$ & $224$ & $220$  \\
2008-07-30  & 03:08:19 & $1800$ & $1.458$ & $2.0$ & $226$ & $230$  \\
2008-07-30  & 03:39:52 & $1800$ & $1.323$ & $2.0$ & $226$ & $224$  \\
2008-07-30  & 04:15:15 & $1800$ & $1.224$ & $2.0$ & $208$ & $216$  \\
2008-07-30  & 04:45:41 & $1800$ & $1.171$ & $2.0$ & $208$ & $206$  \\
2008-07-30  & 05:30:14 & $1800$ & $1.135$ & $2.0$ & $179$ & $187$  \\
2009-04-22  & 08:49:11 & $1800$ & $1.811$ & $2.0$ & $233$ & $235$  \\
2009-04-22  & 09:20:07 & $1800$ & $1.564$ & $2.0$ & $233$ & $232$  \\
2009-04-23  & 08:59:02 & $1800$ & $1.688$ & $2.0$ & $231$ & $234$  \\
2009-04-23  & 09:34:03 & $1800$ & $1.460$ & $2.0$ & $225$ & $230$  \\
2009-04-24  & 09:23:29 & $2400$ & $1.496$ & $2.0$ & $227$ & $231$  \\
\cutinhead{SGAS~J224324.2$-$093508}
2012-08-19  & 02:10:39 & $3600$ & $1.603$ & $1.0$ & $239$ & $240$  \\
2012-08-19  & 03:15:51 & $3600$ & $1.266$ & $1.0$ & $230$ & $234$  \\
2012-08-19  & 04:18:14 & $3600$ & $1.116$ & $1.0$ & $208$ & $220$  \\
2012-08-19  & 05:25:13 & $3600$ & $1.061$ & $1.0$ & $168$ & $186$  \\
\enddata
\tablecomments{Columns are Universal date and time, 
integration time in seconds, sec(z) airmass, 
slit width in arcseconds, 
position angle of the slit (degrees east of north), and 
parallactic angle (degrees east of north). 
Times, airmasses, and parallactic angles are given for the start of each exposure.}
\end{deluxetable*}

%% file: tab_mage_redshifts_v3.tex
\begin{deluxetable*}{lrrrrrrl}
\tabletypesize{\scriptsize}
\tablecolumns{8}
\tablewidth{0pc}
\tablenum{4}
\tablecaption{Redshifts for the \megasaura\ sample.\label{tab:redshifts}}
\tablehead{
\colhead{Source name} &  \colhead{$z_{stars}$} &  \colhead{$\sigma$} & \colhead{$z_{neb}$} & \colhead{$\sigma$} & \colhead{$z_{ISM}$} & \colhead{$\sigma$} & \colhead{Notes}}
\startdata
          SGAS~J000451.7$-$010321 &  1.6812 &  0.0003 &  1.681100 &  0.000100 &  1.680100 &  0.0020 &                                 $z_{neb}$ from Rigby14 \\
          SGAS~J003341.5$+$024217 &  2.3900 &  0.0010 &  2.388770 &  0.000150 &  2.388150 &  0.0003 &        $z_{neb}$ from C~III]; $z_{ism}$ from Si~II~1526 \& Al~II~1670 \\
          SGAS~J010842.2$+$062444 &  1.9099 &  0.0004 &  1.910210 &  0.000030 &  1.907600 &  0.0001 &                                 $z_{neb}$ from Rigby14 \\
   RCSGA~032727$-$132609          &         &         &           &           &           &         &      \\
                        ~~~Knot E &  1.7037 &  0.0004 &  1.703400 &  0.000140 &  1.702300 &  0.0001 &                                 $z_{neb}$ from Rigby14 \\
                        ~~~Knot U &  1.7038 &  0.0007 &  1.703884 &  0.000007 &  1.702400 &  0.0001 &                         $z_{neb}$ from NIRSPEC Rigby15 \\
                        ~~~Knot G &  1.7030 &  0.0010 &  1.703850 &  0.000050 &  1.702500 &  0.0001 &                          $z_{neb}$ from OSIRIS Rigby15 \\
                        ~~~Knot B &     --- &     --- &  1.703600 &  0.000070 &  1.702300 &  0.0002 &                          $z_{neb}$ from OSIRIS Rigby15 \\
                    ~~~counterarc &     --- &     --- &  1.703070 &  0.000060 &  1.702500 &  0.0004 &     $z_{neb}$ from C~III]; $z_{ism}$ from Si~II~1304 \& Al~II~1670 \\
          SGAS~J090003.3$+$223408 &  2.0323 &  0.0005 &  2.032600 &  0.000100 &  2.031700 &  0.0001 &                    $z_{neb}$ from C~III] replaces Rigby15 \\
          SGAS~J095738.7$+$050929 &  1.8205 &  0.0007 &  1.820420 &  0.000040 &  1.820100 &  0.0002 &                                $z_{neb}$ from Rigby+14 \\
          SGAS~J105039.6$+$001730 &     --- &     --- &  3.625300 &  0.000800 &  3.625400 &  0.0006 &                    $z_{ism}$ from Bayliss14; $z_{neb}$ from Bayliss14 \\
              Cosmic Horseshoe SE &  2.3814 &  0.0005 &  2.381150 &  0.000120 &  2.379250 &  0.0003 &                               $z_{neb}$ from Quider+09 \\
          SGAS~J122651.3$+$215220 &  2.9252 &  0.0009 &  2.926000 &  0.000200 &  2.922900 &  0.0001 &                $z_{neb}$ from noisy C~III], replaces Rigby15 \\
          SGAS~J142954.9$+$120239 &  2.8241 &  0.0006 &  2.824500 &       --- &  2.822300 &  0.0001 &                         $z_{neb}$ not well constrained \\
          SGAS~J145836.1$-$002358 &  3.4870 &  0.0010 &  3.486800 &  0.000200 &  3.484950 &  0.0002 &                $z_{neb}$ from FIRE, H$\beta$,[O~III]~5007~\AA \\
          SGAS~J152745.1$+$065219 &         &         &           &           &           &         &      \\
                        ~~~bright &  2.7627 &  0.0007 &  2.762280 &  0.000080 &  2.760650 &  0.0003 &                 $z_{neb}$ from C~III], replaces Rigby15 \\
                    ~~~faint tail &  2.7627 &  0.0007 &  2.762800 &  0.000100 &  2.760100 &  0.0003 &                                  $z_{neb}$ from C~III] \\
          SGAS~J211118.9$-$011431 &  2.8590 &  0.0010 &  2.857700 &       --- &  2.858300 &  0.0007 &                         $z_{neb}$ not well constrained \\
                       Cosmic Eye &  3.0734 &  0.0004 &  3.073500 &  0.000240 &  3.072358 &  0.0003 &  $z_{neb}$ \& $z_{ism}$ from Quider10; multiple $z_{ISM}$ components \\
          SGAS~J224324.2$-$093508 &     --- &     --- &       --- &       --- &  2.080000 &  0.0200 &  Broad-line AGN.  Possible intervening C~IV at $z=2.057$\\
\enddata
\tablecomments{Columns are: 1) source name, 2)--3) stellar redshift and its uncertainty, 4)--5) nebular redshift and its uncertainty, 
6)--7) redshift of the interstellar medium and its uncertainty, and 8) Notes on the source of the redshift measurement.}
\end{deluxetable*}